\documentclass[12pt]{article}
\usepackage{cite}
\usepackage{graphicx}
\usepackage{ amssymb } 
\usepackage{mathtools} 
\pdfoutput=1

\begin{document}

\newcommand{\titulo}{Dimensional crossovers in the\\ Gaussian critical fluctuations above $T_c$\\ of two-layer and three-layer superconductors}

\newcommand{\autor}{A.S.~Viz$^{a}$, M.M.~Botana $^{a}$, J.C.~Verde$^a$, M.V.~Ramallo$^{a,b}$}

\newcommand{\direccion}{$^{a}$Quantum Materials and Photonics Research Group (QMatterPhotonics),\\ Department of Particle Physics, University of Santiago de Compostela,\\ 15782 Santiago de Compostela, Spain \\ \mbox{}\\
$^{b}$Instituto de Materiais (iMATUS),\\ University of Santiago de Compostela, 15782 Santiago de Compostela, Spain}

\begin{center}
  \Large\bf
\titulo\\  \end{center}\mbox{}\vspace{-1cm}\\ 

\begin{center}\normalsize\autor\end{center} 

\begin{center}\normalsize\it\direccion\end{center}


\mbox{}\vskip0cm{\bf Abstract: }

By using a Ginzburg-Landau functional in the Gaussian approximation, we calculate the energy of superconducting  fluctuations above the transition, at zero external magnetic field, of a system composed by a small number {\it N} of parallel two-dimensional superconducting planes, each of them Josephson coupled to its first neighbour, with special focus in the {\it N}=2 and 3 cases. This allows us to obtain expressions for the critical contributions to various observables (fluctuation specific heat and magnetic susceptibility and Aslamazov-Larkin paraconductivity). Our results suggest that these systems may display deviations from pure 2D behaviour and interesting crossover effects, with both similitudes and differences to those known to occur in infinite-layers superconductors. Some challenges for future related research are also outlined.

\mbox{}

{\footnotesize \textbf{Article Highlights:} {\it (i)} We study superconductors composed of a few parallel layers, in the Gaussian-Ginzburg-Landau approach above their critical temperature.
{\it (ii)}  We calculate the heat capacity, susceptibility and conductivity induced by critical thermal fluctuations, mainly for bi- and tri-layers.
{\it (iii)}  We obtain dimensional crossovers in the critical behaviors and compare them with the ones in infinite-layers superconductors.
}

\vfill

\mbox{}\hfill{\footnotesize {\tt mv.ramallo@usc.es}}
\thispagestyle{empty}

\newpage

\newcommand{\abs}[1]{\left\vert#1\right\vert} 
\newcommand{\parent}[1]{\left( #1 \right)}
\newcommand{\corch}[1]{\left[ #1 \right]}
\newcommand{\llave}[1]{\left\{ #1 \right\}}
\newcommand{\unit}[1]{\text{ } \mathrm{#1}}
\newcommand{\function}[2]{#1 \left( #2 \right)}
\newcommand{\prom}[1]{\langle#1\rangle}

\newcommand{\dd}{\mbox{\rm{d}}}
\newcommand{\DF}{\mbox{$\Delta F$}}
\newcommand{\Df}{\mbox{$\Omega$}}
\newcommand{\Ds}{\mbox{$\Delta\sigma$}}

\newcommand{\cfl}{\mbox{$c_{\rm\, fl}$}}
\newcommand{\chifl}{\mbox{$\chi_{\rm\, fl}$}}
\newcommand{\sigmafl}{\mbox{$\sigma_{\rm\, fl  AL}$}}

\newcommand{\ie}{{\it i.e.}}
\newcommand{\eg}{{\it e.g.}}
\newcommand{\etal}{{\it et al. }}

\newcommand{\entrada}[5]{#1 (#5)  #2 #3: #4 }
\newcommand{\inicio}{bollinger,samples1,alegria,katzer}

\setlength{\baselineskip}{18pt}
\


\section{Introduction}\label{sec1}
The different interplays between reduced dimensionality and superconducting properties is a research subject  of  increasing  activity, fostered by the novel posibilities for fabricating nanosized and/or nanostructured superconductors.\cite{\inicio} Also by the fact that  both Cu- and Fe-based high-temperature superconductors are layered materials that may be modelled as stacks of parallel 2D layers.\cite{layer-Fe,ramallovidal} One of the notable effects of low dimensionality  is the enhancement of the critical fluctuations near the superconducting transition temperature $T_c$.\cite{ramallovidal,hohenberg} For instance, it is well known that in 2D films the superconducting fluctuation-induced contributions to various experimental observables above but near $T_c$ are well larger than in 3D bulks. Not only the amplitude, but also the critical exponent is affected.\cite{ramallovidal,hohenberg} For instance, in low-$T_c$ superconductors the fluctuation contribution above $T_c$ to the heat capacity, \cfl, has in 3D bulks critical exponent $x=1/2$ [\ie, $\cfl\propto\varepsilon^{-1/2}$ with $\varepsilon=\ln(T/T_c)$] and in most cases unobservable amplitude,\cite{ramallovidal,hohenberg} while in 2D films the amplitude is well measurable and the critical exponent is $x=1$.\cite{ramallovidal,hohenberg} (For $T<T_c$, fluctuations are also observable in 2D but display the more complex vortex-antivortex phenomenology famously predicted by Kosterlitz and Thouless\cite{KT,HN}). Reduced dimensionality also changes the fluctuation contributions to other observables such as magnetic susceptibility, electrical conductivity, etc.\cite{ramallovidal}

\begin{figure}
\begin{center}\includegraphics[width=0.8\textwidth]{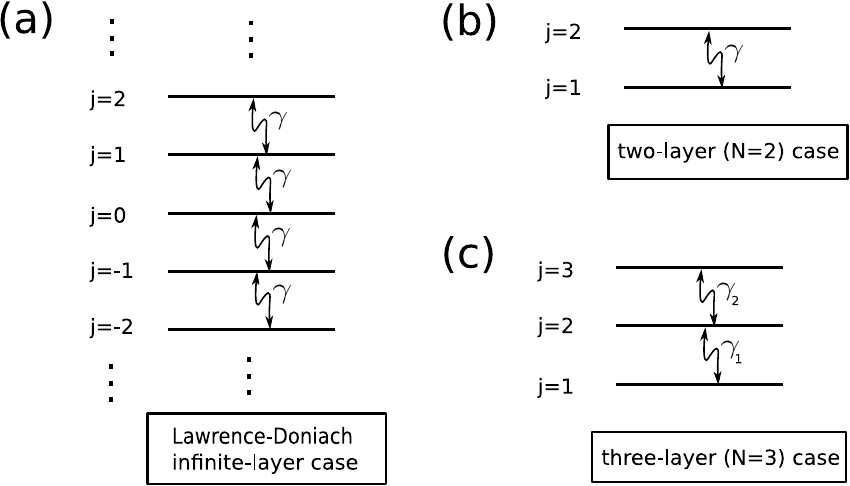}\end{center}\vskip2em
\caption{\label{fig:0} Panel~(a): Schematic representation of a Lawrence-Doniach (LD) or infinite-layer ($N\rightarrow\infty$) superconductor, with single interlayer distances and Josephson couplings between adjacent layers. Panel~(b): Schematic representation of a two-layer superconductor ($N=2$, see Subsection~\ref{sec:2:2} and Section~\ref{sec:discN2}). Panel~(c): Schematic representation of a three-layer superconductor ($N=3$; see Subsection~\ref{sec:2:3}, and also Section~\ref{sec:discN3g} for the $\gamma_1=\gamma_2$ case or Section~\ref{sec:discN3g1g2} for $\gamma_1>\gamma_2$). In~(b) and~(c), each layer $j$ may have a different $T_{cj}$ or a common one; our discussions in Sections~\ref{sec:discN2} to~\ref{sec:discN3g1g2} focus in the latter case.}
\end{figure}

\begin{figure}
\begin{center}\includegraphics[width=0.5\textwidth]{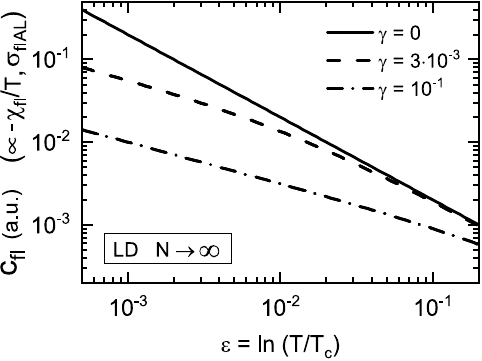}\end{center}
\caption{\label{fig:1}Fluctuation specific heat \cfl\ from the well-known GGL-LD predictions for superconductors composed of an infinite number of parallel 2D planes, as a function of the reduced temperature $\varepsilon$ and for different values of the Josephson-coupling constant $\gamma$ between adjacent layers. (As a reference, for  optimally-doped cuprates of the YBaCuO family values $\gamma\simeq0.001\sim0.05$ are usually proposed{\cite{ vina, carballeira, rey}}). The \cfl\ is given in arbitrary units, and is proportional to the also observables $-\chifl/T$ and \sigmafl\ (see main text for details). The figure illustrates that when $\varepsilon\ll\gamma$ the \cfl\ behaves as in a 3D system (somewhat decreased amplitude and log-log slope -1/2) while if $\varepsilon\gg\gamma$ it displays a 2D behaviour (log-log slope -1). See also Fig.~\ref{fig:2}.}
\end{figure}

Some of the richer phenomenologies for the interrelations between low dimensionality and critical fluctuations are provided by layered superconductors. These may be modelled using the Lawrence-Doniach (LD) functional,\cite{LD} \ie, the Ginzburg-Landau (GL) free energy  for a superconductor composed of an infinite (macroscopic) number of parallel planes, each of them Josephson-coupled with its adjacent neighbour. Panel (a) of Figure~\ref{fig:0} schematizes such superconductors. By introducing small (Gaussian) excitations, it is possible to calculate expressions for the critical fluctuations above $T_c$\cite{LD} that are in good agreement with measurements  in various macroscopic layered materials, including for instance the Cu-and Fe-based high-$T_c$ superconductors.\cite{vina,carballeira,rey} The basic prediction of this LD modelization for the fluctuation-induced heat capacity above $T_c$ under zero external magnetic field \cfl\ may be written as:\cite{ramallovidal,LD,tsuzuki,quader}
\begin{equation}
  \cfl=\frac{A_{\rm TF}}{\varepsilon}\left(1+\frac{B_{\rm LD}}{\varepsilon}\right)^{-1/2},
  \label{eq-LD}
\end{equation}
where $A_{\rm TF}=k_{\rm B}/[4\pi\xi_{ab}^2(0)s]$ is the Thouless-Ferrell amplitude,\cite{thouless,ferrell} $B_{\rm LD}=[2\xi_{c}(0)/s]^2$ is the LD parameter,\cite{ramallovidal,LD,tsuzuki,quader} $s$ is the inter-layer distance and $\xi_{ab}(0)$ and $\xi_{c}(0)$ are the GL amplitudes of the coherence length in the in-plane and out-of-plane directions. The latter is given in terms of the Josephson coupling constant between adjacent planes, $\gamma$, as $\xi_{c}(0)=s\sqrt{\gamma}$.\cite{ramallovidal,LD} A representation of the resulting \cfl\  is given in Fig.~\ref{fig:1}. We also plot, in Fig.~\ref{fig:2}, the corresponding critical exponent (calculated as the log-log slope of the \cfl-vs-$\varepsilon$ curve) showing that it crosses over the 2D ($x=1$) and 3D ($x=1/2$) values as $\varepsilon$ decreases and $T_c$ is approached [and as the inter-plane correlation grows by $\xi_c(\varepsilon)=\xi_c(0)\varepsilon^{-1/2}$; note that the crossover is located at around  $\varepsilon_{\rm crossover}\simeq B_{\rm LD}=4\gamma$.] 

The LD calculations have been generalized by various authors to a number of different cases, including for instance non-Gaussian fluctuations,{\cite{dorsey,lang1}} inclusion of high-temperature effects,{\cite{lang1,vidal,prl,mosqueira}} or also considering  an infinite amount of layers but  with two alternating interlayer Josephson  couplings $\gamma_1$ and $\gamma_2$.{\cite{ramallovidal,klemm,buzdin}}

However,  to our knowledge the critical fluctuations in superconductors composed of only a few layers [see Figure~\ref{fig:0}(b) and~(c)] have not been calculated yet, even in the relatively simple Gaussian-Ginzburg-Landau (GGL) approximation above $T_c$. This will be the  main purpose of the present work, with a focus on identifying possible dimensional crossover effects due to the Josephson couplings.

Let us note that a topic with some mathematical resemblance may be multi-band superconductors (with each band corresponding to the gap in different sheets of the Fermi surface) when Josephson-like expresions are chosen for the interband coupling. This case was considered in terms of  fluctuations, \eg, in~{\cite{REF1a}}. However, this is a different physical problem in various respects, the main ones being that such couplings do not introduce dimensional crossovers\cite{REF1a} (consequently with the fact that they do not correspond to spatial variations of the gaps) and that the interband interactions change $T_c$ differently to the few-layer case.{\cite{REF1b,REF1c,REF4a}}

In the present  article, we consider a GL functional of such a few-layer system and calculate the effects of critical fluctuations near but above the critical temperature, in the GGL approximation, for some of the main observables in the zero-external magnetic field limit (fluctuation heat capacity, magnetic suscetibility and electrical conductivity). We find explicit expression for various cases, and  physically discuss the dimensional crossover effects induced by the inter-layer Josephson couplings  in such geometries, focusing mainly in the two-layer and three-layer cases. Our results  suggest that the finite-layer superconductors have the capability to display dimensional crossover effects quite comparable, in the variety of its phenomenology, to those in the LD model for infinite-layers superconductors. This includes, for instance, deviations from the 2D values of the critical exponents or crossovers  of the amplitudes of the fluctuations when $\varepsilon$, and hence $T/T_c$, varies.

We organize this  article  as follows: In Sect.~\ref{sec:Basic} we write our basic equations and calculate the GGL fluctuation spectra. In Sect.~\ref{sec:Expressions} we write the resulting fluctuation contributions to three observables (the fluctuation specific heat, \cfl, the fluctuation-induced magnetic susceptibility, \chifl, and the  Aslamazov-Larkin electrical paraconductivity $\sigmafl$); we also write expressions for their corresponding critical exponents, $x$, and amplitudes, the latter through a so-called effective number of independent fluctuating planes $N_e$ that will be helpful for the interpretation of the results. In Sect.~\ref{sec:discN2} we discuss these results for two-layer superconductors in terms of their $x$ and $N_e$ crossovers as $\varepsilon$ varies, for different Josephson couplings. In  Sects.~\ref{sec:discN3g} and~\ref{sec:discN3g1g2} we  discuss the three-layer superconductors, for different values of  the Josephson couplings and their ratio.  In Sect.~\ref{sec:Conclusions} we summarize our conclusions and briefly comment on some of the difficulties and challenges for further reseach  in this topic.

\section{\label{sec:Basic}Basic expressions for the Gaussian-Ginzburg-Landau fluctuations above $T_c$ in two-layer and three-layer superconductors}

\subsection{GL functional}

As starting point, let us model a superconductor composed by a (small) number $N$ of parallel superconducting planes, each of them Josephson-coupled to its adjacent neighbour, by writing its Ginzburg-Landau (GL) functional as composed of the sum of free energies intrinsic to each $j =1, \dots N$ plane,  plus interactions between each $j$ and $j+1$ planes:
\begin{equation}
	\label{eq:DF}
	\DF = \sum_{j=1}^{N}  { \DF_j^{\rm intr} +\sum_{j=1}^{N-1}{\DF_{j,j+1}^{\rm inter}   }},
\end{equation}
with 
\begin{equation}
	\label{eq:DFintr}
	\DF_j^{{\rm intr}} = a_0\; \int \dd^2 \mbox{\bf r}\;  \left\{ \varepsilon_j \lvert\psi_j\rvert^2 + \frac{b}{2a_0}  \lvert\psi_j\rvert^4 +  \xi^2_{ab}(0)  \rvert\nabla_{xy} \psi_j \rvert^2 \right\},
\end{equation}
where $\psi_j$ are the GL wavefunctions of each plane, $\mbox{\bf r}=(x,y)$ are the in-plane coordinates, and $a_0$, $b$ and $\xi_{ab}(0)$  are  the GL constants and the  in-plane coherence length amplitude. Also, $\varepsilon_j$ is the reduced temperature of each plane:
\begin{equation}
	\label{eq:epsilon}
	\varepsilon_j = \ln \left( {T}/{T_{cj}} \right) \simeq (T-T_{cj})/T_{cj},
\end{equation}
where $T_{cj}$ is its intrinsic critical temperature. In these initial equations we consider the general case in which $T_{cj}$ and $\varepsilon_{j}$ may be different in each plane, but let us note already here that  many of our discussions in the present article will focus, for concreteness, in the case in which all critical temperatures coincide ($T_{cj}=T_{c}$, and hence also $\varepsilon_{j}=\varepsilon$, for all $j$). 
We also emphasize that we used the same $\xi_{ab}(0)$ for all the planes, which is probably a fair approximation if all of them are of the same material. Let us note that for most superconductors eventual variations of $\xi_{ab}(0)$ are linked to variations of $T_c$ of greater extent, so that we expect that any eventual effects due to $\xi_{ab}(0)$ variations between layers are expected to be smaller than the corresponding effects due to different $T_{cj}.$

For the inter-plane interaction term between planes $j$ and $j+1$ we employ (as is also done by the usual LD functional for infinite-layers superconductors):
\begin{equation}
	\DF_{j,j+1}^{\rm inter} = a_0 \int \dd^2 \mbox{\bf r}  \;\left\{ \; \gamma_j  \,\lvert \psi_j - \psi_{j+1} \rvert^2\;\right\},
	\label{eq:DFi}
\end{equation}
where $\gamma_j$ is a Josephson coupling constant between the planes $j$ and $j+1$. Note that in the limit $N\rightarrow\infty$ our functional given by Eqs.~\refeq{eq:DF} to \refeq{eq:DFi} simply recovers the usual LD functional for infinite-layers superconductors.\cite{LD,LD2}  Note also that we assumed, in Eqs.~\refeq{eq:DFintr} and \refeq{eq:DFi}, zero external magnetic field and negligible effects of the potential vector gauge field (the latter would be important for the  Kosterlitz-Thouless fluctuations below $T_c$). This is because we focus in this paper on  $H=0$ and for  temperatures  sufficiently above $T_c$ as to be in the Gaussian-Ginzburg-Landau (GGL) region of the fluctuations. In that region, the $\lvert \psi \rvert^4 $ term in Eq.~\refeq{eq:DFintr} may be neglected and independent fluctuation modes, and their corresponding free energy,  are searched. In the rest of this article we proceed with that program for $N=2$ and 3, and discuss the results. 

\subsection{\label{sec:2:2}GGL fluctuation modes for $N=2$}

In the $N=2$ case, we have two (potentially different) intrinsic $T_c$'s, and hence two reduced temperatures $\varepsilon_1$ and $\varepsilon_2$, and only one interlayer Josephson coupling constant $\gamma_1 = \gamma$.  When considering this $N=2$ case, the Eqs.~\refeq{eq:DF} to \refeq{eq:DFi} in the GGL approximation above $T_c$ may be rewritten in explicit matrix form as:
\begin{equation}
	\label{eq:DF-matrix}
	\Df (\psi_1,\psi_2)= \left( \begin{array}{cc}
		\psi_1^* & \psi_2^* \end{array} \right) \left( \begin{array}{cc}
		\varepsilon_1 + \gamma & -\gamma  \\
		-\gamma & \varepsilon_2+\gamma  \end{array} \right) \left( \begin{array}{c}
		\psi_1 \\
		\psi_2 \end{array} \right),
\end{equation}
where $\Df$ is an interlayer contribution so that the total GL functional is:
\begin{equation}
\DF = a_0 \sum_{\alpha=Re,Im}\int \dd^2 \mbox{\bf r} \left\{ \xi^2_{ab}(0) \sum_{j=1,2} \lvert \nabla_{xy} \psi_j^{\alpha} \rvert^2 + \Df (\psi_1^{\alpha},\psi_2^{\alpha})\right\}.
\label{eq:DF-W-real}
\end{equation}
In this expression it has been convenient to separate the wavefunctions into their real and imaginary parts, labeled by the  index $\alpha$. Note that \refeq{eq:DF-W-real} can be also written in $k_{xy}$-Fourier space as: 
\begin{equation}
\DF \propto \sum_{\alpha=Re,Im}\int \dd k_x \dd k_y \llave{\xi^2_{ab}(0) k_{xy}^2\sum_{j=1,2}\abs{\psi_{jk_{xy}}^{\alpha}}^2 + \Df(\psi_{1k_{xy}}^{\alpha},\psi_{2k_{xy}}^{\alpha})}.
\label{eq:DF-W-momentos}
\end{equation}

We now diagonalize the $2 \times 2$ matrix in \refeq{eq:DF-matrix}. This leads to 
\begin{equation}
	\Df (\psi_1,\psi_2)= \left( \begin{array}{cc}
		f_1^* & f_2^* \end{array} \right) \left( \begin{array}{cc}
		\omega_1 & 0  \\
		0 & \omega_2  \end{array} \right) \left( \begin{array}{c}
		f_1 \\
		f_2 \end{array} \right),
\end{equation}
with
\begin{equation}
	\omega_1= \frac{1}{2} \corch{\varepsilon_1 + \varepsilon_2 + 2\gamma - \sqrt{(\varepsilon_1-\varepsilon_2)^2 + 4 \gamma^2}},
\end{equation}
\begin{equation}
	\omega_2= \frac{1}{2} \corch{\varepsilon_1 + \varepsilon_2 + 2\gamma + \sqrt{(\varepsilon_1-\varepsilon_2)^2 + 4 \gamma^2}}.
\end{equation}
The $f_{1,2}$ themselves are linear combinations of $\psi_{1,2}$ that in this small-$N$ case may be expressed in a relatively compact form:
\begin{equation}
	\label{eq:fs1}
f_1 = \frac{\left(\omega_2-\varepsilon_1-\gamma\right)\psi_1 +\gamma\psi_2}{    \sqrt{\left(\omega_2-\varepsilon_1-\gamma\right)^2+\gamma^2}},
\end{equation}
\begin{equation}
\label{eq:fs2}
f_2 = \frac{\left(\omega_1-\varepsilon_1-\gamma\right)\psi_1 +\gamma\psi_2}{    \sqrt{\left(\omega_1-\varepsilon_1-\gamma\right)^2+\gamma^2}},
\end{equation}
Note that in the limit  of zero Josephson interplane coupling these quotients become simpler: In particular,  for $\gamma \rightarrow 0$  it is $f_1\rightarrow  \psi_1$ and $f_2  \rightarrow \psi_2$ when $\varepsilon_2>\varepsilon_1$, or   $f_1  \rightarrow \psi_2$ and  $f_2  \rightarrow -\psi_1$ when $\varepsilon_1>\varepsilon_2$ (see next paragraph for the case $\varepsilon_1=\varepsilon_2$; we used l'H\^opital's rule for the simultaneous zeroes in the  numerator and  denominator of Eqs.~\refeq{eq:fs1} and~\refeq{eq:fs2}).

\subsubsection*{The case $N=2$ with $T_{c1}=T_{c2}\; (=T_c)$} 
Let us here consider $N=2$ but with  all the planes having the same critical temperature, and hence also $\varepsilon_1=\varepsilon_2=\varepsilon$. In that case, the inter-layer GGL energy eigenvalues $\omega_1$ and $\omega_2$ become:
\begin{equation}
    \omega_1 = \varepsilon,
    \label{omega1N2}
\end{equation}
\begin{equation}
    \omega_2 = \varepsilon + 2\gamma,
    \label{omega2N2}
\end{equation}
and the $f_1$, $f_2$ eigenwavefunctions are:
\begin{equation}
    f_1 = (\psi_1 + \psi_2)/\sqrt{2},
\end{equation}
\begin{equation}
    f_2 = (\psi_2 - \psi_1)/\sqrt{2}.
\end{equation}

\subsection{\label{sec:2:3}GGL fluctuation modes for $N=3$}

For $N=3$, the matrix form of \Df\ becomes:
\begin{equation}
	\Df (\psi_1,\psi_2,\psi_3)=  \left( \begin{array}{ccc}
		\psi_1^* & \psi_2^* & \psi_3^* \end{array} \right) \left( \begin{array}{ccc}
		\varepsilon_1 + \gamma_1 & -\gamma_1 & 0  \\
		-\gamma_1 & \varepsilon_2+\gamma_1 +\gamma_2 & - \gamma_2 \\
		0 & -\gamma_2 &  \varepsilon_3 + \gamma_2 \end{array} \right) \left( \begin{array}{c}
		\psi_1 \\
		\psi_2 \\
		\psi_3 \end{array} \right).
\end{equation}
Diagonalizing this matrix is possible with the use of Cardano's formulas for the roots of third order polynomials.  The expression of the corresponding eigenvalues $\omega_1$ to $\omega_3$ are considerably long and therefore we devote Appendix~\ref{Ap1} to write them. In the next subsection, we consider a more manegeable case. 

\subsubsection*{The case $N=3$ with $T_{c1}=T_{c2}=T_{c3}\; (=T_c)$} 
Fortunately, the cumbersome general $N=3$ expressions for $\omega_{1,2,3}$  dramatically collapse in size when considering the case in which all the planes share the same critical temperature. In this case, the inter-layer GGL energy eigenvalues simply become:
\begin{equation}
    \omega_1 = \varepsilon,
    \label{omega1N3}
\end{equation}
\begin{equation}
    \omega_2 = \varepsilon + \gamma_1+\gamma_2-\sqrt{\gamma_1^2-\gamma_1\gamma_2+\gamma_2^2},
     \label{omega2N3}
\end{equation}
\begin{equation}
    \omega_3 = \varepsilon + \gamma_1+\gamma_2+\sqrt{\gamma_1^2-\gamma_1\gamma_2+\gamma_2^2},
     \label{omega3N3}
\end{equation}
where again $\varepsilon=\varepsilon_1=\varepsilon_2=\varepsilon_3$.

\subsection{The quantity $\sum_{j=1}^N \omega_j^{-1}$}

From such $\omega$ eigenvalues of the GGL functional, in principle most fluctuation-induced observables quantities may be obtained.  In this regard, of particular significance will be the quantity $\sum \omega_j^{-1}$ because it  will be proportional, in the GGL approach above $T_c$, to the fluctuation-induced heat capacity \cfl\ (see next Section; it will be also proportional to $-\chifl/T$ and $\sigmafl$).\cite{ramallovidal,klemm}

In the $N=2$ case with a single $T_c$, this quantity becomes:
\begin{equation}
    \sum_{j=1}^2 \omega_j^{-1} = \frac{1}{\varepsilon}\;
    \frac{2\varepsilon+2\gamma}
    {\varepsilon+2\gamma}.
    \label{IN2}
\end{equation}

In the $N=3$ case with a single $T_c$, it becomes:
\begin{equation}
    \sum_{j=1}^3 \omega_j^{-1} = \frac{1}{\varepsilon}\;
    \frac{3\varepsilon^2+3\gamma_1\gamma_2+4\varepsilon(\gamma_1+\gamma_2)}
    {\varepsilon^2+3\gamma_1\gamma_2+2\varepsilon(\gamma_1+\gamma_2)}.
    \label{IN3}
\end{equation}

\section{\label{sec:Expressions}Fluctuation-induced heat capacity, magnetic susceptibility and AL paraconductivity}

\subsection{Expressions for \cfl, \chifl\ and \sigmafl}
From the GGL free energy written in terms of independent modes, it is possible to calculate its thermal statistical averages and then the fluctuation-induced contributions to various observables. In particular, for the basic averages of the independent modes, as expected it is   $\prom{f_{jkxy}^{\alpha\;2}} \propto k_{\rm B} T/[\xi_{ab}^2(0)k_{xy}^2+\omega_{j}]$, where not only the inter-plane contribution appears but also the in-plane kinetic energy term.\footnote{In our expressions  $k_{\rm B}, \mu_0, \phi_0, e$ and $\hbar$ are the usual universal physical constants} This is very similar to the case in the LD model, except for the substitution of the LD spectrum $\omega_{k_z}^{\rm LD}=2\gamma(1-\cos k_z s)$ by our $\omega_j$. Therefore it is easy to adapt to our case  well-known LD calculations for the superconducting fluctuation contributions to various observables. In particular, for the following ones (always considered above $T_c$ and in the limit of zero external magnetic field):

For the fluctuation-induced specific heat, \cfl\ (see, \eg, \cite{ramallovidal,klemm} for  a parallel calculation in the  LD case):
\begin{equation}
\label{eq:cflbase}
\cfl=\frac{k_{\rm B}}{4\pi\xi_{ab}^2(0)L_z}   
\;\sum_{j=1}^{N}\omega_j^{-1},
\end{equation}
where $L_z$ is the thickness of the $N$-layer system. 

For the fluctuation-induced magnetic susceptibility, \chifl, with the magnetic field perpendicular to the layers and always in the weak magnetic field limit (see, \eg, \cite{ramallovidal} for a similar calculations in the LD case):
\begin{equation}
\frac{-\chifl}{T}=\frac{\mu_0\pi k_{\rm B}\xi_{ab}^2(0)}{3\phi_0^2L_z}
\;\sum_{j=1}^{N}\omega_j^{-1}.
\end{equation}

For the in-plane electrical conductivity, we also calculated  (adapting the procedures of \cite{ramallovidal,hohenberg,klemm})  the  direct fluctuation contribution  (also known as Aslamazov-Larkin paraconductivity $\sigmafl${), that is} the dominant contribution to  the experimental $\sigma_{\rm fl}$ at least in high-temperature cuprates{\cite{vina,carballeira,rey}}:\footnote{For \sigmafl, we are assuming a sample with enough distance between electrical contacts for the inter-plane {\it resistance} to be well smaller than the in-plane one, so that all layers must be averaged in the conduction. This is generally the case expected in experiments with real few-layer films.}
\begin{equation}
\sigmafl=\frac{e^2}{16\hbar L_z}
\;\sum_{j=1}^{N}\omega_j^{-1}.
\end{equation}

When combined with our explicit formulae  for the quantity $\sum\omega_j^{-1}$ for $N=2$ and $N=3$ (Eqs.~\refeq{IN2} and \refeq{IN3}), the above  expressions for \cfl,  \chifl\ and  \sigmafl\ become also explicit.

\subsection{Critical exponents}

In order to physically discuss the above results for \cfl, \chifl\ and \sigmafl, a first quantity of interest will be the critical exponent, defined as the log-log slope of the plot of the fluctuation heat capacity versus reduced-temperature:
\begin{equation}
	x= -\frac{\partial \ln \cfl}{\partial \ln \varepsilon}.
	\label{eq:defx}
\end{equation}
Note that the same critical exponent is going to be shared with $-\chifl/T$ and $\sigmafl$. Note also that in the 2D limit it is $\cfl\propto \varepsilon^{-1}$ and therefore $x=1$ (while  for 3D bulks it is $x=1/2$). 

When applied to the Eqs.~\refeq{IN2} to ~\refeq{eq:cflbase} obtained in the previous sections, Eq.~\refeq{eq:defx} leads to the following result for the $N=2$ case:
\begin{equation}
x=
\frac{\varepsilon^2+2\gamma\varepsilon+2\gamma^2}
{   \left(\varepsilon+ \gamma\right)\;\left(\varepsilon+ 2 \gamma\right)   },
\end{equation}
and for the $N=3$ case:
\begin{equation}
x=
\frac{
3 \varepsilon^4
+8  (\gamma_1+\gamma_2)\varepsilon^3
+8  (\gamma_1+\gamma_2)^2\varepsilon^2
+12  \gamma_1 \gamma_2 (\gamma_1+\gamma_2)\varepsilon
+9 \gamma_1^2 \gamma_2^2
}
{\left[
\varepsilon^2+3 \gamma_1 \gamma_2+2 \varepsilon (\gamma_1+\gamma_2)
\right]
\;\left[
3 \varepsilon^2+3 \gamma_1 \gamma_2+4 \varepsilon (\gamma_1+\gamma_2)
\right]}.
\end{equation}

Both of these expressions saturate to the pure 2D value $x=1$ in the limit of zero Josephson coupling between planes ($\gamma_j\rightarrow 0$), as it could be expected.

\subsection{Effective number of independent fluctuating planes}

We also introduce now a second relevant quantity, informing about the amplitude of the fluctuactions.  We shall call this the ``effective number of independently fluctuating superconducting planes'', $N_e$, and we define it as 
\begin{equation}
	N_e= \frac{\cfl}{\cfl^{N=1}} = \frac{\chi_{\rm fl}}{\chi_{\rm fl}^{N=1}} = \frac{\sigma_{\rm flAL}}{\sigma_{\rm flAL}^{N=1}}.
\end{equation}
In other words, this quantity is the increment of the fluctuactions with respect to the value expected for a $N=1$ 2D layer (with the same $L_z$).  A value $N_e=1$ would indicate all of the $N$ planes are fluctuating together as a single plane, and is expected to correspond at least to the limit $\gamma \rightarrow \infty$ (strong inter-plane correlation).  In contrast, a value $N_e =N$ is expected to be recovered at least in the limit $\gamma \rightarrow 0$ (no inter-plane correlations, each plane fluctuates independently of the {other}).

\section{\label{sec:discN2}Discussion of the results for two-layer superconductors}

Let us now present a more physical discussion of the  consequences of the expressions obtained up to now, starting here with the simpler $N=2$ case (we defer $N=3$ to the  Sects.~\ref{sec:discN3g} and~\ref{sec:discN3g1g2}). 

We first note that in this $N=2$ case the interlayer fluctuation energy is split into two contributions, the ones of Eqs.~\refeq{omega1N2} and \refeq{omega2N2}, what may be understood as one half of the fluctuation modes having the same  energy as in a regular 2D layer, and the other half having the fluctuation energy of a 2D layer but  with an ``effective'' reduced temperature $\varepsilon + 2\gamma$ [or with effective critical temperature $T_c/\exp{(2\gamma)}$]. Logically, the total fluctuation superfluid density accumulates both contributions, and so does $\cfl^{N=2}$ (via the quantity $\sum\omega_j^{-1}$). 

In the case $\gamma = 0$ (no interlayer interactions) both independent modes behave with the same effective critical temperature. In that case, as it could be expected the critical exponent is the 2D value, $x=1$, and the effective number of independently fluctuating planes becomes $N_e=2$:
\begin{equation}
    N_e (\gamma = 0)  \;=\;  2,
\end{equation}
\begin{equation}
    x(\gamma = 0)  \;=\;  1.
\end{equation}

In the case with $\gamma \rightarrow \infty$ we expect however the two planes acting as a single one, and in fact in that limit we get:
\begin{equation}
    N_e (\gamma \rightarrow  \infty)  \;=\;  1,
\end{equation}
\begin{equation}
    x (\gamma \rightarrow \infty)  \;=\;  1.
\end{equation}
Note that the limit $\gamma \rightarrow \infty$ has the physical meaning that any variation of the superconducting wave function between adjacent layers would be energetically prohibitive, so that the only physically relevant situation in the statistical averages would be having the two layers acting as a single one - what directly should imply $x=1$ and $N_e=1$, as the above equations confirm. (These equations can be also understood by considering that if $\gamma \rightarrow \infty$ the $f_2$ fluctuating mode becomes too difficult to excite and does not contribute to \cfl).

\begin{figure}
\begin{center}\includegraphics[width=0.5\textwidth]{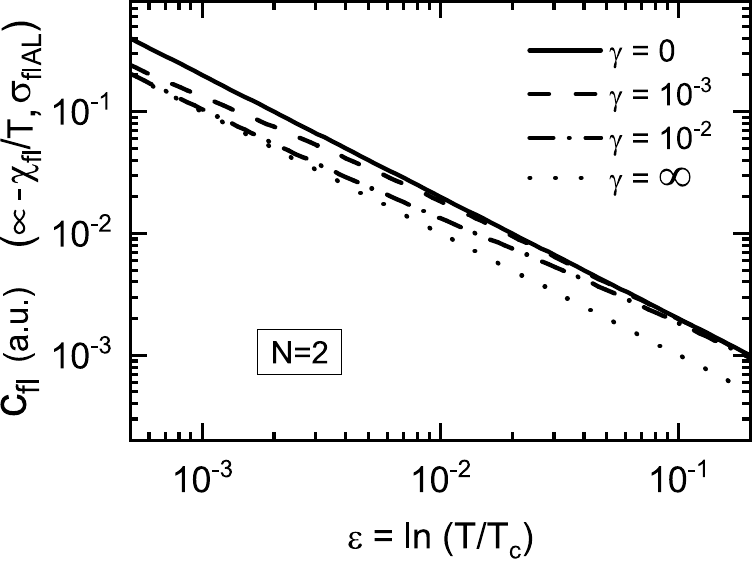}\end{center}
\caption{\label{fig:3}Fluctuation specific heat \cfl\ from our expressions for two-layer superconductors,  as a function of the reduced temperature $\varepsilon$ and for different values of the Josephson coupling  $\gamma$. The \cfl\ is given in arbitrary units (and is proportional to the also observables $-\chifl/T$ and \sigmafl).  See also Figs.~\ref{fig:4} and~\ref{fig:5} for an interpretation of the results.}
\end{figure}

Between these two pure 2D limits ($x=1$ with either $N_e = 1$ or $2$) intermediate cases must appear, in which   the inter-plane correlations will result in deviations of the critical exponent from the 2D value, $x\neq1$.  Also, $N_e$ must undergo a crossover between $N_e = 2$ and $N_e = 1$ as $\gamma$ evolves from $0$ to $\infty$. This is represented in Figs.~\ref{fig:3} to~\ref{fig:5}. In particular,  Fig.~\ref{fig:3} plots $N_e ^{N=2}$ versus  $\varepsilon$ for different values of the inter-layer coupling $\gamma$. For $\gamma \rightarrow \infty$ and $\gamma = 0$, the limiting values 1 and 2 are obtained, as commented before. For intermediate   $\gamma$ values, also $N_e \rightarrow 1$ if $\varepsilon \rightarrow 0$. This agrees with the fact that when $\varepsilon \rightarrow 0$ both planes are expected to be strongly correlated due to the growth (and divergence at $T=T_c$) of the coherence length between them (that may be estimated as $\xi_c(\varepsilon) = \xi_c(0) / \sqrt{\varepsilon}$ with $\xi_c(0)=L_z/\sqrt{\gamma}$, in analogy to the usual LD model for infinite-layers superconductors). In contrast, as $\varepsilon\rightarrow \infty$ and $\xi_c(\varepsilon) \rightarrow 0$ both planes will become progressively independent and $N_e=N$ ($=2$ in this case), as confirmed by Fig.~\ref{fig:3}. A rough estimate of the midpoint of this $N_e$  crossover may be obtained from   $\xi_c(\varepsilon) \sim L_z$, again in analogy to what occurs in the LD model. This leads to $\varepsilon_{\rm crossover} \sim \gamma$,  in good agreement with Fig.~\ref{fig:3}. 

In Fig.~\ref{fig:4}, it may be observed a phenomenology for the critical exponent $x$ that is consequent with the above considerations. In particular, for both $\gamma = 0$ and $\gamma \rightarrow \infty$ a pure 2D value $x = 1$ is obtained (irrespectively of $N_e=1$ or $2$ the system behaves as a planar one). This pure 2D exponent is also obtained for intermediate values of $\gamma$ when either $\varepsilon \rightarrow 0$ or $\varepsilon \rightarrow \infty$, corresponding to the fact that also  $N_e \rightarrow 1$ or $2$. But when both $\gamma$ and $\varepsilon$ have intermediate values, a deviation from the pure 2D behaviour appears, indicating precursor correlations in the third dimension. Then, $x$ becomes intermediate between the 2D and 3D  values ($x=1$ and $1/2$). In fact, the minimum of $x(\varepsilon)$, calculable by $\partial x / \partial \varepsilon = 0$, just happens at $\varepsilon_{\rm crossover} =  \sqrt{2} \,\gamma$, corresponding to $x \approx 0.83$, which is similar to what was estimated above  for the $N_e$ crossover. Therefore, we conclude that this $N=2$ finite layer case has a  capability of displaying intermediate-dimensionality crossover not very far from what happens in the  infinite-layers case, although without the capability of reaching the 3D limit.

\section{\label{sec:discN3g}Discussion of the results for three-layer superconductors  with $\gamma_1 = \gamma_2 \;(=\gamma)$} 

We now explore the physical consequences of the expressions obtained for the $N = 3$ case. For concreteness, we first consider the case in which $\gamma_1$ and $\gamma_2$ take a common value $\gamma$ (in the Sect.~\ref{sec:discN3g1g2}   we shall consider the $\gamma_1\neq\gamma_2$ case). As in the $N = 2$ case, the relevant quantities will be $\cfl$, $N_e$ and $x$.

First of all,  note that when $\gamma = 0$ we obtain the expected 2D result ($x=1$), with $N_e = 3$ as also expected (each plane behaves independently and acts twodimensionally):
\begin{equation}
     N_e (\gamma_1 = \gamma_2 = 0) \;=\; 3,
\end{equation}
\begin{equation}
    x (\gamma_1 = \gamma_2 = 0) \;=\; 1.
\end{equation}

The equations also reproduce the expected result for the opposite limit $\gamma \rightarrow \infty$, in which the three planes should act as a single one. In that {case, the equations} produce $x=1$ and $N_e=1$ as it corresponds to that physical situation:
\begin{equation}
     N_e (\gamma_1 = \gamma_2 \rightarrow \infty)  \;=\;  1,
\end{equation}
\begin{equation}
    x (\gamma_1 = \gamma_2 \rightarrow \infty)  \;=\;  1.
\end{equation}

\begin{figure}
\begin{center}\includegraphics[width=0.5\textwidth]{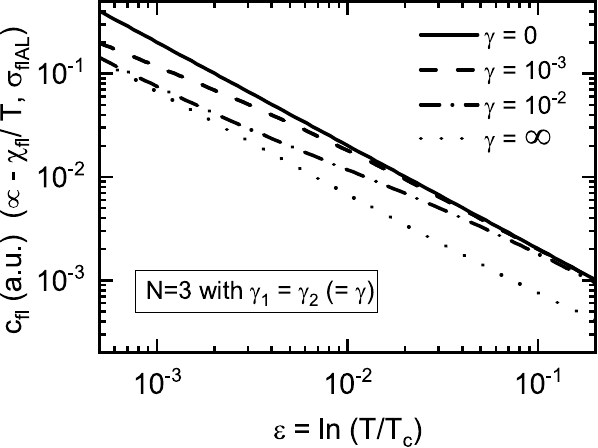}\end{center}
\caption{\label{fig:6}Fluctuation specific heat \cfl\ from our expressions for three-layer superconductors with a single Josephson coupling,  $\gamma=\gamma_1=\gamma_2$, as a function of the reduced temperature $\varepsilon$ and for different  $\gamma$. The \cfl\ is given in arbitrary units (and is proportional to the also observables $-\chifl/T$ and \sigmafl).  See also Figs.~\ref{fig:7} and~\ref{fig:8} for an interpretation of the results.}
\end{figure}

\begin{figure}
\begin{center}\includegraphics[width=0.5\textwidth]{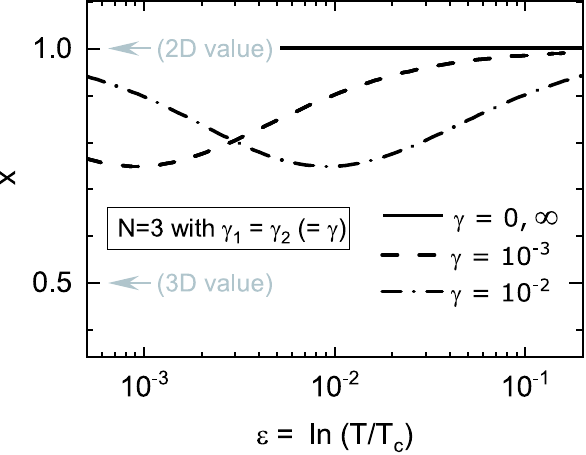}\end{center}
\caption{\label{fig:7}Critical exponent $x$ of \cfl\ (and of $-\chifl/T$ and $\sigmafl$) for three-layer superconductors with a single Josephson coupling,  $\gamma=\gamma_1=\gamma_2$,  as a function of $\varepsilon$ and for different  $\gamma$. The figure illustrates  deviations from the 2D value ($x=1$) when  $\varepsilon$ and $\gamma$ take comparable values, which may be further understood when contrasted with the $N_e$ evolution in Fig.~\ref{fig:8}.}
\end{figure}

\begin{figure}
\begin{center}\includegraphics[width=0.5\textwidth]{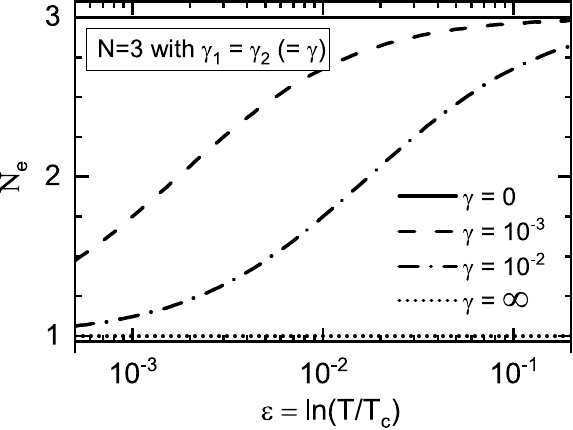}\end{center}
\caption{\label{fig:8}Effective number of independent fluctuating planes, $N_e$, for three-layer superconductors with a single Josephson coupling,  $\gamma=\gamma_1=\gamma_2$,  as a function of  $\varepsilon$ and for different    $\gamma$. The figure illustrates crossovers between the $N_e=1$ and $3$ values (with no plateau at $N_e\simeq2$) as  $\varepsilon$ and $\gamma$ vary (and with them the corresponding inter-plane  correlations) correlated with the evolutions of the critical exponent $x$ in Fig.~\ref{fig:7}.}
\end{figure}

Between these two pure 2D limits, intermediate-dimensionality cases must appear for intermediate values of $\gamma$. This is represented in Figs.~\ref{fig:6} to~\ref{fig:8}. In Fig.~\ref{fig:8}, $N_e$ is plotted versus  $\varepsilon$ for different values of $\gamma$. As expected, there is a crossover as $\varepsilon$ increases from $N_e = 1$ up to $N_e = 3$. The crossover temperature increases as $\gamma$ decreases (and for $\gamma = 0$ or $\gamma \rightarrow \infty$  the crossover is outside of the experimental window). In Fig.~\ref{fig:7} the critical exponent $x$ is plotted versus  reduced temperature. Again, the $x(\varepsilon)$ behaviour is correlated with the evolution of $N_e$: When $N_e = 1$ or $3$, $x$ takes the 2D value $x=1$, and when $N_e$ is crossing over those values the system develops a non-2D critical exponent (becoming closer to the 3D value the further away $N_e$ is from its limiting values $1$ or $3$).

\section{\label{sec:discN3g1g2}Discussion of the results for  three-layer superconductors   with $\gamma_1 > \gamma_2$} 

We now explore the case $N=3$ with  significantly different interlayer Josephson couplings $\gamma_1$ and $\gamma_2$. For concreteness, we take  $\gamma_1/\gamma_2 >1$ (but note that the equations are symmetrical  to  interchanges of $\gamma_1$ and $\gamma_2$). 

\begin{figure}
\begin{center}
\includegraphics[width=0.45\textwidth]{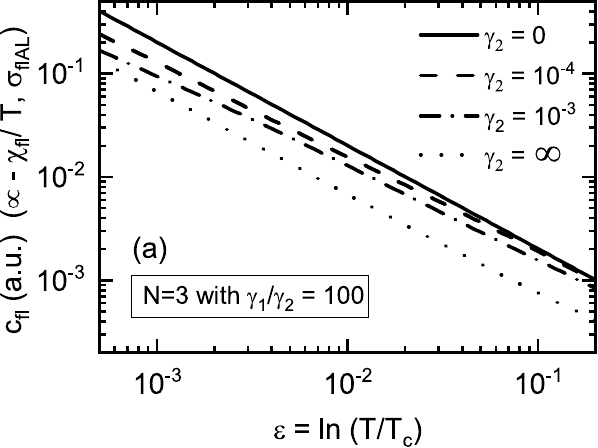}\hspace{3em}\includegraphics[width=0.45\textwidth]{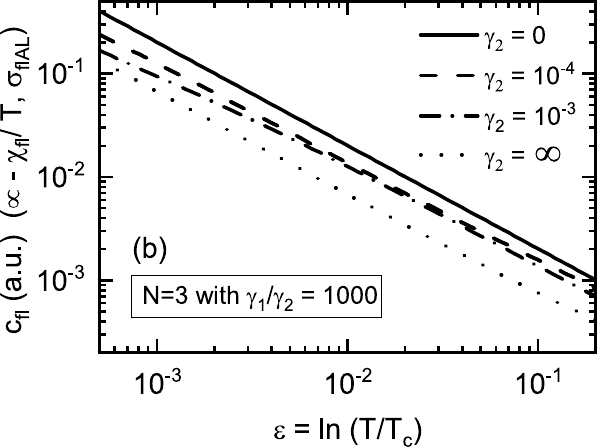}\end{center}
\caption{\label{fig:9}Panel (a): Fluctuation specific heat \cfl\ from our expressions for three-layer superconductors with different Josephson couplings  $\gamma_1/\gamma_2=100$, as a function of the reduced temperature $\varepsilon$ and for different values of $\gamma_2$.  Panel (b): Same for an increased $\gamma_1/\gamma_2=1000$. See also Figs.~\ref{fig:10} and~\ref{fig:11}  for an interpretation of the results.}
\end{figure}

\begin{figure}
\begin{center}
\includegraphics[width=0.45\textwidth]{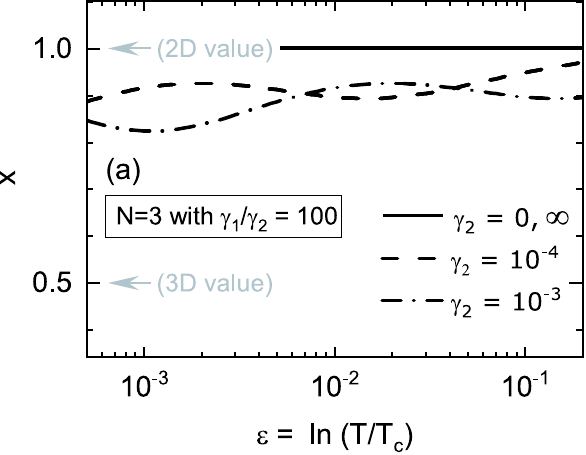}\hspace{3em}\includegraphics[width=0.45\textwidth]{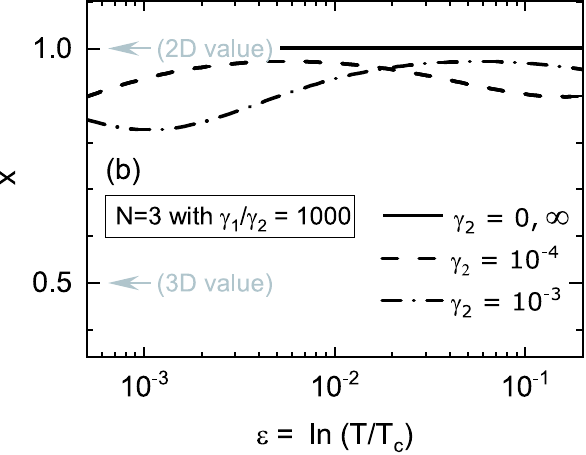}\end{center}
\caption{\label{fig:10}Panel (a): Critical exponent $x$ of \cfl\  for three-layer superconductors with  $\gamma_1/\gamma_2=100$,  as a function of $\varepsilon$ and for different  $\gamma_2$. The figure hints at double-featured  deviations from the 2D value ($x=1$)  which may be correlated with the $N_e$ changes (and plateaus) in Fig.~\ref{fig:11} (see also main text). Panel (b): Same for an increased $\gamma_1/\gamma_2=1000$, illustrating a softening (rather than a displacement) of the $x\neq1$ features.}
\end{figure}

\begin{figure}
\begin{center}
\includegraphics[width=0.45\textwidth]{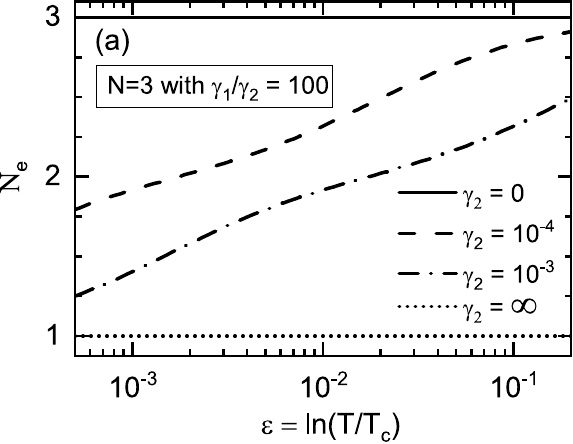}\hspace{3em}\includegraphics[width=0.45\textwidth]{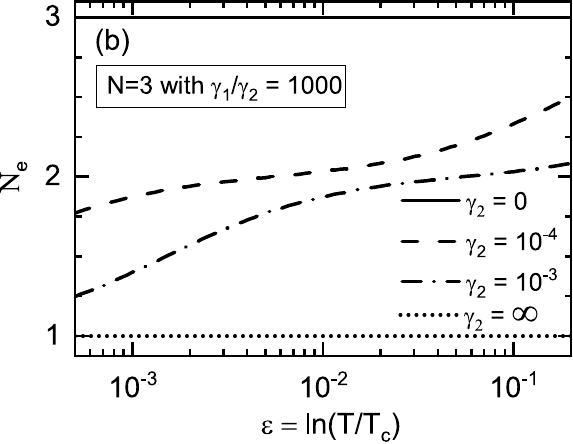}\end{center}
\caption{\label{fig:11}Panel (a): Effective number of independent fluctuating planes, $N_e$,  for three-layer superconductors with  $\gamma_1/\gamma_2=100$,  as a function of $\varepsilon$ and for different  $\gamma_2$. The figure illustrates not only a crossover between  $N_e=1$ and $3$, but also a small plateau around $N_e\simeq2$. Panel (b): Same for  $\gamma_1/\gamma_2=1000$, demonstrating an enlargement of the plateau around $N_e\simeq2$ (correlated to the softening of the $x\neq1$ features in Fig.~\ref{fig:10}).}
\end{figure}

Figs.~\ref{fig:9} to~\ref{fig:11} [panels (a) for $\gamma_1/\gamma_2 = 100$ and panels (b) for $\gamma_1/\gamma_2 = 1000$] display the $\cfl$, $x$ and $N_e$ versus  reduced temperature obtained for $N = 3$ and different values of $\gamma_2$. 

As it happened in the previous Section, for  $\gamma_2 = 0$ and $\gamma_2 \rightarrow \infty$ two different 2D limit cases are obtained, with $x=1$ and $N_e = N = 3$ for $\gamma_2 = 0 $, and with  $x=1$ and $N_e = 1$ for $\gamma_1 \rightarrow \infty$. 

Intermediate dimensionality behaviour appears for intermediate values of $\gamma_2$ (and hence $\gamma_1$), in which $x$ may develop deviations from the 2D value simultaneously to deviations of $N_e$ from its saturation values $1$ or $3$. But an interesting additional feature may appear at certain reduced temperatures, in which $N_e$ plateaus at $N_e = 2$. This must correspond to the case in which two of the layers have already saturated their mutual correlation, while the third still develops fluctuations not locked to the ones of the other layers. This feature may be seen in our Figs.~\ref{fig:9} to~\ref{fig:11}, mainly in those corresponding to $\gamma_1/\gamma_2 = 1000$ [\ie, panels (b)], while for the lower $\gamma_1/\gamma_2 = 100$ the $N_e$ plateau is well smaller. The resulting evolution of $x(\varepsilon)$ becomes then of non-trivial aspect (also for $\gamma_1/\gamma_2 = 100$), though it may be understood as a  double valey of deepness tracking the slope of $N_e(\varepsilon)$. We conclude therefore that this $N = 3$ case not only is able to display intermediate dimensionality behaviour in comparable significance to the infinite-layers  case, but also that this case is to some extent able to display richer phenomenology (multiple crossovers) in spite of never reaching true 3D ($x=1/2$) behaviour.

\section{\label{sec:Conclusions}Conclusions and some remaining challenges}
\begin{figure}
\begin{center}\includegraphics[width=0.5\textwidth]{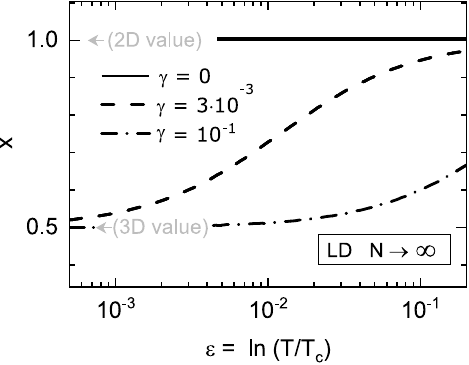}\end{center}
\caption{\label{fig:2}Critical exponent $x$ of \cfl\ (and of $-\chifl/T$ and $\sigmafl$) resulting from  the  GGL-LD calculations for infinite-layers superconductors, as a function of the reduced temperature $\varepsilon$ for different values of the Josephson coupling  $\gamma$. The figure illustrates the crossover from the 3D value ($x=1/2$) to the 2D one ($x=1$) as $\varepsilon$ evolves from  $\varepsilon\ll\gamma$ to $\varepsilon\gg\gamma$, ant that  the dimensional corossover occurrs  around $\varepsilon_{\rm crossover}\simeq 4\gamma$.}
\end{figure}

\begin{figure}
\begin{center}\includegraphics[width=0.5\textwidth]{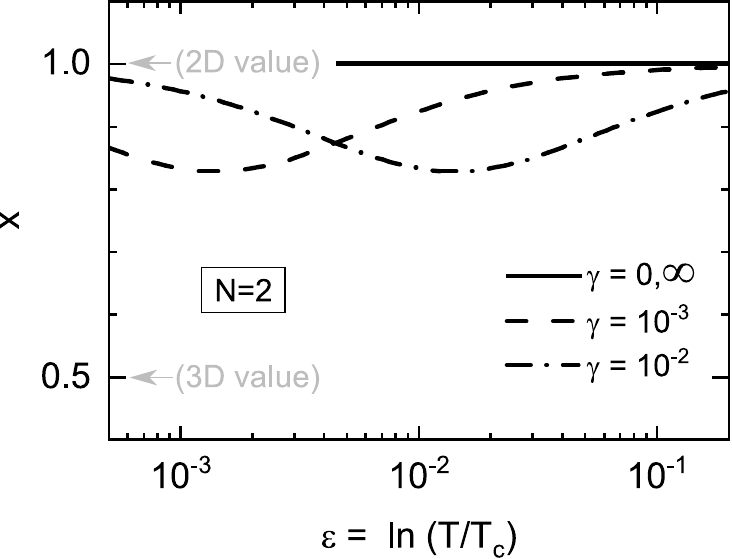}\end{center}
\caption{\label{fig:4}Critical exponent $x$ of \cfl\ (and of $-\chifl/T$ and $\sigmafl$)  for two-layer superconductors,  as a function of $\varepsilon$ and for different  $\gamma$. The figure illustrates  deviations from the 2D value ($x=1$) when  $\varepsilon$ and $\gamma$ take comparable values, which may be further understood when contrasted with the $N_e$ evolution in Fig.~\ref{fig:5}.}
\end{figure}

\begin{figure}
\begin{center}\includegraphics[width=0.5\textwidth]{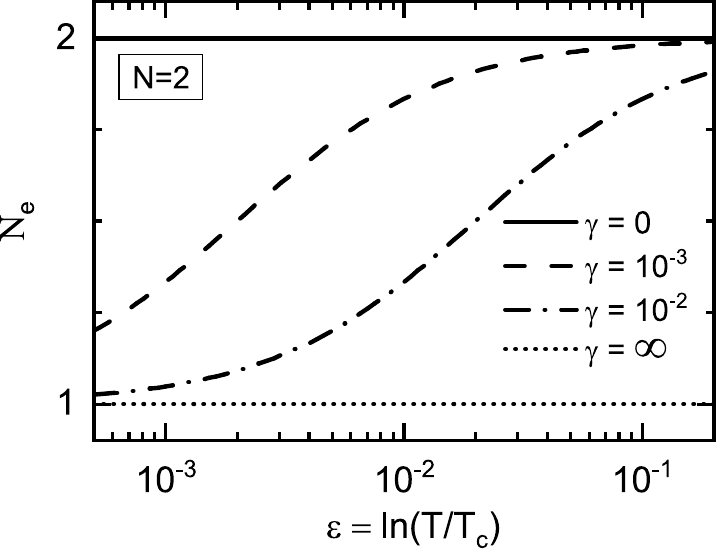}\end{center}
\caption{\label{fig:5}Effective number of independent fluctuating planes, $N_e$,  for two-layer superconductors,  as a function of  $\varepsilon$ and for different    $\gamma$. The figure illustrates crossovers between the $N_e=1$ and $2$ values as  $\varepsilon$ and $\gamma$ vary, and with them the corresponding inter-plane  correlations. These crossovers may be correlated with the evolutions of the critical exponent $x$ in Fig.~\ref{fig:4}.}
\end{figure}

In conclusion, we have considered a GL functional of a few-layer superconductor (mainly two- and tree-layer) and calculated the effects of critical fluctuations  above the critical temperature, in the GGL approximation, for some of the main observables in the zero-external magnetic field limit (fluctuation heat capacity, magnetic suscetibility and electrical conductivity). The resulting expressions suggest the capability of these systems to display  crossover effects on the critical exponents and amplitudes, with similitudes and differences with respect to  those predicted by the Lawrence-Doniach (LD) model for infinite-layers superconductors. For instance, in the bi-layer ($N=2$) case the critical exponent develops deviations from the pure 2D value as the temperature approaches $T_c$ (as in the LD model) but, instead of crossing over from the 2D to the 3D values (see Fig.~\ref{fig:2}), it undergoes a different evolution (see Fig.~\ref{fig:4}) of critical spatial dimensionality  (2D-intermediate dimensionality-2D), including two 2D regimes with different effective number of independently fluctuating planes (see Fig.~\ref{fig:5}). Also, for $N=3$ the evolution of the critical exponent displays a similar frustrated change of dimensionality plus an evolution of the number of independent layers from $N_e=1$ to $N_e=3$.

Let us finally briefly comment on some of the expected challenges and difficulties on further studying these potentially interesting superconducting fluctuations of few-layer systems. First, in spite of sample availability now being far easier than in the {past,\cite{\inicio}} the specimens are bound to be small in volume; this could make measurements of the heat capacity challenging, probably favouring magnetic screening or electrical measurements (and hence $\chi$ and/or  $\sigma$). Smallness also makes boundary conditions more important, and while for negligible external magnetic fields and above the transition (the case studied in this article) a change in the value of $T_c$ may be expected to roughly summarize most of these  boundary effects, for other situations involving well-developed vortices (sizeable magnetic fields, temperatures below the transtion, etc.) the constraints imposed by the substrate of the sample will have to be taken into account, both experimentally and theoretically. Also, because of these substrate effects and other issues, it could be important to further extend our  calculations  to the case with different $T_c$ for each plane, only hinted at in the present article. Probably more challenging may be to extend them to the case with larger number of planes, as the difficulty of the matrix diagonalization increases considerably with $N$, what could constraint  calculations to be only numerical instead of analytical.

{\bf Acknowledgments: } \quad
 This work was supported by the Agencia Estatal de Investigaci\'on (AEI) and Fondo Europeo de Desarrollo Regional (FEDER) through project PID2019-104296GB-100, by Xunta de Galicia (grant GRC number ED431C 2018/11) and iMATUS (2021 internal project RL3). JCV was supported by
the Spanish Ministry of Education for grant FPU14/00838. MMB was supported by Ministerio de Universidades of Spain through the National Program FPU (gran number FPU19/05266).


\clearpage

\clearpage

\clearpage

\clearpage

\appendix

\section{Appendix: Full expression for $\omega_j$ in the $N=3$ case}\label{Ap1}
The following are the complete expressions for the fluctuation energy spectrum for $N=3$ and arbitrary $\varepsilon_1$, $\varepsilon_2$, $\varepsilon_3$, $\gamma_1$ and $\gamma_2$:

\mbox{}

\tiny

\noindent\begin{equation*}
\omega_1 = 1/3 (\varepsilon_1 + \varepsilon_2 + \varepsilon_3 + 2 \gamma_1 + 2 \gamma_2) - (2^{
     1/3} (-(-\varepsilon_1 - \varepsilon_2 - \varepsilon_3 - 2 \gamma_1 - 2 \gamma_2)^2 + 
       3 (\varepsilon_1 \varepsilon_2 +
\end{equation*}
\begin{equation*}
       \varepsilon_1 \varepsilon_3 + \varepsilon_2 \varepsilon_3 + \varepsilon_1 \gamma_1 + \varepsilon_2 \gamma_1 + 
          2 \varepsilon_3 \gamma_1 + 2 \varepsilon_1 \gamma_2 + \varepsilon_2 \gamma_2 + \varepsilon_3 \gamma_2 + 
          3 \gamma_1 \gamma_2)))/(3 (2 \varepsilon_1^3 - 3 \varepsilon_1^2 \varepsilon_2 - 
       3 \varepsilon_1 \varepsilon_2^2 + 2 \varepsilon_2^3 - 3 \varepsilon_1^2 \varepsilon_3 +
\end{equation*}
\begin{equation*} 
12 \varepsilon_1 \varepsilon_2 \varepsilon_3 - 
       3 \varepsilon_2^2 \varepsilon_3 - 3 \varepsilon_1 \varepsilon_3^2 - 3 \varepsilon_2 \varepsilon_3^2 + 2 \varepsilon_3^3 + 
       3 \varepsilon_1^2 \gamma_1 - 12 \varepsilon_1 \varepsilon_2 \gamma_1 + 3 \varepsilon_2^2 \gamma_1 + 
       6 \varepsilon_1 \varepsilon_3 \gamma_1 + 6 \varepsilon_2 \varepsilon_3 \gamma_1 - 6 \varepsilon_3^2 \gamma_1 + 
       6 \varepsilon_1 \gamma_1^2+
\end{equation*}
\begin{equation*}
 6 \varepsilon_2 \gamma_1^2 - 12 \varepsilon_3 \gamma_1^2 + 16 \gamma_1^3 - 
       6 \varepsilon_1^2 \gamma_2 + 6 \varepsilon_1 \varepsilon_2 \gamma_2 + 3 \varepsilon_2^2 \gamma_2 + 
       6 \varepsilon_1 \varepsilon_3 \gamma_2 - 12 \varepsilon_2 \varepsilon_3 \gamma_2 + 3 \varepsilon_3^2 \gamma_2 - 
       6 \varepsilon_1 \gamma_1 \gamma_2 + 12 \varepsilon_2 \gamma_1 \gamma_2 - 6 \varepsilon_3 \gamma_1 \gamma_2 - 
       \end{equation*}
\begin{equation*}
       6 \gamma_1^2 \gamma_2 - 12 \varepsilon_1 \gamma_2^2 + 6 \varepsilon_2 \gamma_2^2 + 
       6 \varepsilon_3 \gamma_2^2 - 6 \gamma_1 \gamma_2^2 + 
       16 \gamma_2^3 + 
        ((2 \varepsilon_1^3 - 3 \varepsilon_1^2 \varepsilon_2 - 
            3 \varepsilon_1 \varepsilon_2^2 + 2 \varepsilon_2^3 - 3 \varepsilon_1^2 \varepsilon_3 + 
            12 \varepsilon_1 \varepsilon_2 \varepsilon_3 - 3 \varepsilon_2^2 \varepsilon_3 - 3 \varepsilon_1 \varepsilon_3^2 - 
            3 \varepsilon_2 \varepsilon_3^2 + 
\end{equation*}
\begin{equation*}
2 \varepsilon_3^3 + 3 \varepsilon_1^2 \gamma_1  - 
            12 \varepsilon_1 \varepsilon_2 \gamma_1 + 3 \varepsilon_2^2 \gamma_1 + 6 \varepsilon_1 \varepsilon_3 \gamma_1 + 
            6 \varepsilon_2 \varepsilon_3 \gamma_1 - 6 \varepsilon_3^2 \gamma_1 + 6 \varepsilon_1 \gamma_1^2 + 
            6 \varepsilon_2 \gamma_1^2 - 12 \varepsilon_3 \gamma_1^2 + 16 \gamma_1^3 - 
            6 \varepsilon_1^2 \gamma_2 + 
            \end{equation*}
\begin{equation*}
6 \varepsilon_1 \varepsilon_2 \gamma_2 + 3 \varepsilon_2^2 \gamma_2 + 
            6 \varepsilon_1 \varepsilon_3 \gamma_2 - 12 \varepsilon_2 \varepsilon_3 \gamma_2 + 3 \varepsilon_3^2 \gamma_2 - 
            6 \varepsilon_1 \gamma_1 \gamma_2 + 12 \varepsilon_2 \gamma_1 \gamma_2 - 6 \varepsilon_3 \gamma_1 \gamma_2 - 
            6 \gamma_1^2 \gamma_2 - 12 \varepsilon_1 \gamma_2^2 + 6 \varepsilon_2 \gamma_2^2 + 
            \end{equation*}
\begin{equation*}
            6 \varepsilon_3 \gamma_2^2 - 6 \gamma_1 \gamma_2^2 + 16 \gamma_2^3)^2 + 
          4 (-(-\varepsilon_1 - \varepsilon_2 - \varepsilon_3 - 2 \gamma_1 - 2 \gamma_2)^2 + 
             3 (\varepsilon_1 \varepsilon_2 + \varepsilon_1 \varepsilon_3 + \varepsilon_2 \varepsilon_3 + \varepsilon_1 \gamma_1 + 
                \varepsilon_2 \gamma_1 + 2 \varepsilon_3 \gamma_1 + 2 \varepsilon_1 \gamma_2 + 
            \end{equation*}
\begin{equation*}
                \varepsilon_2 \gamma_2 + 
                \varepsilon_3 \gamma_2 + 3 \gamma_1 \gamma_2))^3)^{1/2})^{1/3}) + (1/(
  3 2^{1/3}))((2 \varepsilon_1^3 - 3 \varepsilon_1^2 \varepsilon_2 - 3 \varepsilon_1 \varepsilon_2^2 + 2 \varepsilon_2^3 -
    3 \varepsilon_1^2 \varepsilon_3 + 12 \varepsilon_1 \varepsilon_2 \varepsilon_3 - 3 \varepsilon_2^2 \varepsilon_3 - 
    3 \varepsilon_1 \varepsilon_3^2 -
    \end{equation*}
\begin{equation*}
    3 \varepsilon_2 \varepsilon_3^2 + 2 \varepsilon_3^3 + 3 \varepsilon_1^2 \gamma_1 - 
    12 \varepsilon_1 \varepsilon_2 \gamma_1 + 3 \varepsilon_2^2 \gamma_1 + 6 \varepsilon_1 \varepsilon_3 \gamma_1 + 
    6 \varepsilon_2 \varepsilon_3 \gamma_1 - 6 \varepsilon_3^2 \gamma_1 + 6 \varepsilon_1 \gamma_1^2 + 
    6 \varepsilon_2 \gamma_1^2 - 12 \varepsilon_3 \gamma_1^2 + 16 \gamma_1^3 -
    \end{equation*}
\begin{equation*}
6 \varepsilon_1^2 \gamma_2 + 
    6 \varepsilon_1 \varepsilon_2 \gamma_2 + 3 \varepsilon_2^2 \gamma_2 + 6 \varepsilon_1 \varepsilon_3 \gamma_2 - 
    12 \varepsilon_2 \varepsilon_3 \gamma_2 + 3 \varepsilon_3^2 \gamma_2 - 6 \varepsilon_1 \gamma_1 \gamma_2 + 
    12 \varepsilon_2 \gamma_1 \gamma_2 - 6 \varepsilon_3 \gamma_1 \gamma_2 - 6 \gamma_1^2 \gamma_2 - 
    \end{equation*}
\begin{equation*}
    12 \varepsilon_1 \gamma_2^2 + 6 \varepsilon_2 \gamma_2^2 + 6 \varepsilon_3 \gamma_2^2 - 6 \gamma_1 \gamma_2^2 + 
    16 \gamma_2^3 +
    ((2 \varepsilon_1^3 - 3 \varepsilon_1^2 \varepsilon_2 - 3 \varepsilon_1 \varepsilon_2^2 + 
         2 \varepsilon_2^3 - 3 \varepsilon_1^2 \varepsilon_3 + 12 \varepsilon_1 \varepsilon_2 \varepsilon_3 - 
         3 \varepsilon_2^2 \varepsilon_3 - 3 \varepsilon_1 \varepsilon_3^2 - 3 \varepsilon_2 \varepsilon_3^2 + 2 \varepsilon_3^3 +
         \end{equation*}
\begin{equation*}
         3 \varepsilon_1^2 \gamma_1 - 12 \varepsilon_1 \varepsilon_2 \gamma_1 + 3 \varepsilon_2^2 \gamma_1 + 
         6 \varepsilon_1 \varepsilon_3 \gamma_1 + 6 \varepsilon_2 \varepsilon_3 \gamma_1 - 6 \varepsilon_3^2 \gamma_1 +
         6 \varepsilon_1 \gamma_1^2 + 6 \varepsilon_2 \gamma_1^2 - 12 \varepsilon_3 \gamma_1^2 + 16 \gamma_1^3 - 
         6 \varepsilon_1^2 \gamma_2 + 6 \varepsilon_1 \varepsilon_2 \gamma_2 + 3 \varepsilon_2^2 \gamma_2 + 
         \end{equation*}
\begin{equation*}
         6 \varepsilon_1 \varepsilon_3 \gamma_2 - 12 \varepsilon_2 \varepsilon_3 \gamma_2 + 3 \varepsilon_3^2 \gamma_2 - 
         6 \varepsilon_1 \gamma_1 \gamma_2 + 12 \varepsilon_2 \gamma_1 \gamma_2 - 6 \varepsilon_3 \gamma_1 \gamma_2 - 
         6 \gamma_1^2 \gamma_2 - 12 \varepsilon_1 \gamma_2^2 + 6 \varepsilon_2 \gamma_2^2 + 
         6 \varepsilon_3 \gamma_2^2 - 6 \gamma_1 \gamma_2^2 + 16 \gamma_2^3)^2 + 
         \end{equation*}
\begin{equation*}
       4 (-(-\varepsilon_1 - \varepsilon_2 - \varepsilon_3 - 2 \gamma_1 - 2 \gamma_2)^2 + 
          3 (\varepsilon_1 \varepsilon_2 + \varepsilon_1 \varepsilon_3 + \varepsilon_2 \varepsilon_3 + \varepsilon_1 \gamma_1 + \varepsilon_2 \gamma_1 + 2 \varepsilon_3 \gamma_1 + 2 \varepsilon_1 \gamma_2 + \varepsilon_2 \gamma_2 +
             \varepsilon_3 \gamma_2 + 3 \gamma_1 \gamma_2))^3)^{1/2} )^{1/3})
\end{equation*}

\mbox{}

\mbox{}

\noindent\begin{equation*}
\omega_2 = 1/3 (\varepsilon_1 + \varepsilon_2 + \varepsilon_3 + 2 \gamma_1 + 
     2 \gamma_2) + ((1 + 
       i \sqrt{3}) (-(-\varepsilon_1 - \varepsilon_2 - \varepsilon_3 - 2 \gamma_1 - 2 \gamma_2)^2 + 
       3 (\varepsilon_1 \varepsilon_2 + \varepsilon_1 \varepsilon_3 + \varepsilon_2 \varepsilon_3 +
              \end{equation*}              
              \begin{equation*}
 \varepsilon_1 \gamma_1 + \varepsilon_2 \gamma_1 + 
          2 \varepsilon_3 \gamma_1 + 2 \varepsilon_1 \gamma_2 + \varepsilon_2 \gamma_2 + \varepsilon_3 \gamma_2 + 
          3 \gamma_1 \gamma_2)))/(3 2^{
     2/3} (2 \varepsilon_1^3 - 3 \varepsilon_1^2 \varepsilon_2 - 3 \varepsilon_1 \varepsilon_2^2 + 2 \varepsilon_2^3 - 
       3 \varepsilon_1^2 \varepsilon_3 + 12 \varepsilon_1 \varepsilon_2 \varepsilon_3 - 
              \end{equation*}
\begin{equation*}       
       3 \varepsilon_2^2 \varepsilon_3 - 
       3 \varepsilon_1 \varepsilon_3^2 - 3 \varepsilon_2 \varepsilon_3^2 + 2 \varepsilon_3^3 + 3 \varepsilon_1^2 \gamma_1 -
       12 \varepsilon_1 \varepsilon_2 \gamma_1 + 3 \varepsilon_2^2 \gamma_1 + 6 \varepsilon_1 \varepsilon_3 \gamma_1 +
       6 \varepsilon_2 \varepsilon_3 \gamma_1 - 6 \varepsilon_3^2 \gamma_1 +
6 \varepsilon_1 \gamma_1^2 + 
                            \end{equation*}
\begin{equation*} 
       6 \varepsilon_2 \gamma_1^2 - 12 \varepsilon_3 \gamma_1^2 + 16 \gamma_1^3 - 6 \varepsilon_1^2 \gamma_2 + 
       6 \varepsilon_1 \varepsilon_2 \gamma_2 + 3 \varepsilon_2^2 \gamma_2 + 6 \varepsilon_1 \varepsilon_3 \gamma_2 - 
       12 \varepsilon_2 \varepsilon_3 \gamma_2 + 3 \varepsilon_3^2 \gamma_2 - 6 \varepsilon_1 \gamma_1 \gamma_2 + 
       12 \varepsilon_2 \gamma_1 \gamma_2 -
                                          \end{equation*}
\begin{equation*} 
6 \varepsilon_3 \gamma_1 \gamma_2 -6 \gamma_1^2 \gamma_2 - 
       12 \varepsilon_1 \gamma_2^2 + 6 \varepsilon_2 \gamma_2^2 + 6 \varepsilon_3 \gamma_2^2 - 
       6 \gamma_1 \gamma_2^2 + 
       16 \gamma_2^3 + ((2 \varepsilon_1^3 - 3 \varepsilon_1^2 \varepsilon_2 - 
            3 \varepsilon_1 \varepsilon_2^2 + 2 \varepsilon_2^3 - 3 \varepsilon_1^2 \varepsilon_3 + 
            12 \varepsilon_1 \varepsilon_2 \varepsilon_3 -
 \end{equation*}
\begin{equation*}           
            3 \varepsilon_2^2 \varepsilon_3 -3 \varepsilon_1 \varepsilon_3^2-
            3 \varepsilon_2 \varepsilon_3^2 + 
2 \varepsilon_3^3 + 3 \varepsilon_1^2 \gamma_1 - 
            12 \varepsilon_1 \varepsilon_2 \gamma_1 + 3 \varepsilon_2^2 \gamma_1 + 6 \varepsilon_1 \varepsilon_3 \gamma_1 + 
            6 \varepsilon_2 \varepsilon_3 \gamma_1 - 6 \varepsilon_3^2 \gamma_1 + 6 \varepsilon_1 \gamma_1^2 + 
            6 \varepsilon_2 \gamma_1^2 -
            \end{equation*}
\begin{equation*} 
            12 \varepsilon_3 \gamma_1^2 + 16 \gamma_1^3 - 
            6 \varepsilon_1^2 \gamma_2 + 6 \varepsilon_1 \varepsilon_2 \gamma_2 + 3 \varepsilon_2^2 \gamma_2 + 
            6 \varepsilon_1 \varepsilon_3 \gamma_2 - 12 \varepsilon_2 \varepsilon_3 \gamma_2 + 3 \varepsilon_3^2 \gamma_2 -
            6 \varepsilon_1 \gamma_1 \gamma_2 + 
                        \end{equation*}
\begin{equation*} 
12 \varepsilon_2 \gamma_1 \gamma_2 - 6 \varepsilon_3 \gamma_1 \gamma_2 - 
            6 \gamma_1^2 \gamma_2 - 12 \varepsilon_1 \gamma_2^2 + 6 \varepsilon_2 \gamma_2^2 + 
            6 \varepsilon_3 \gamma_2^2 - 6 \gamma_1 \gamma_2^2 + 16 \gamma_2^3)^2 + 
          4 (-(-\varepsilon_1 - \varepsilon_2 - \varepsilon_3 - 2 \gamma_1 - 2 \gamma_2)^2 + 
            \end{equation*}
\begin{equation*} 
             3 (\varepsilon_1 \varepsilon_2 + \varepsilon_1 \varepsilon_3 + \varepsilon_2 \varepsilon_3 + \varepsilon_1 \gamma_1 + 
                \varepsilon_2 \gamma_1 + 2 \varepsilon_3 \gamma_1 + 2 \varepsilon_1 \gamma_2 + \varepsilon_2 \gamma_2 + 
                \varepsilon_3 \gamma_2 + 3 \gamma_1 \gamma_2))^3)^{1/2})^{1/3}) - (1/(
  6\cdot 2^{1/3}))(1 - 
  \end{equation*}
\begin{equation*} 
i \sqrt{3}) (2 \varepsilon_1^3 - 3 \varepsilon_1^2 \varepsilon_2 - 
     3 \varepsilon_1 \varepsilon_2^2 + 2 \varepsilon_2^3 - 3 \varepsilon_1^2 \varepsilon_3 + 12 \varepsilon_1 \varepsilon_2 \varepsilon_3 - 
     3 \varepsilon_2^2 \varepsilon_3 - 3 \varepsilon_1 \varepsilon_3^2 - 3 \varepsilon_2 \varepsilon_3^2 + 2 \varepsilon_3^3 +  3 \varepsilon_1^2 \gamma_1 - 12 \varepsilon_1 \varepsilon_2 \gamma_1 + 
     \end{equation*}
\begin{equation*} 
3 \varepsilon_2^2 \gamma_1 + 
     6 \varepsilon_1 \varepsilon_3 \gamma_1 + 6 \varepsilon_2 \varepsilon_3 \gamma_1 - 6 \varepsilon_3^2 \gamma_1 + 
     6 \varepsilon_1 \gamma_1^2 + 6 \varepsilon_2 \gamma_1^2 - 12 \varepsilon_3 \gamma_1^2 + 16 \gamma_1^3 - 
     6 \varepsilon_1^2 \gamma_2 + 6 \varepsilon_1 \varepsilon_2 \gamma_2 + 3 \varepsilon_2^2 \gamma_2 + 
     \end{equation*}
\begin{equation*}
     6 \varepsilon_1 \varepsilon_3 \gamma_2 - 12 \varepsilon_2 \varepsilon_3 \gamma_2 + 3 \varepsilon_3^2 \gamma_2 - 
     6 \varepsilon_1 \gamma_1 \gamma_2 + 12 \varepsilon_2 \gamma_1 \gamma_2 - 6 \varepsilon_3 \gamma_1 \gamma_2 -
     6 \gamma_1^2 \gamma_2 - 12 \varepsilon_1 \gamma_2^2 + 6 \varepsilon_2 \gamma_2^2 + 6 \varepsilon_3 \gamma_2^2 - 
     6 \gamma_1 \gamma_2^2 + 
         \end{equation*}
\begin{equation*}
     16 \gamma_2^3 + ((2 \varepsilon_1^3 - 3 \varepsilon_1^2 \varepsilon_2 - 3 \varepsilon_1 \varepsilon_2^2 + 
          2 \varepsilon_2^3 - 3 \varepsilon_1^2 \varepsilon_3 + 12 \varepsilon_1 \varepsilon_2 \varepsilon_3 - 
          3 \varepsilon_2^2 \varepsilon_3 - 3 \varepsilon_1 \varepsilon_3^2 - 3 \varepsilon_2 \varepsilon_3^2 + 2 \varepsilon_3^3 + 
     \end{equation*}
\begin{equation*}
          3 \varepsilon_1^2 \gamma_1 - 12 \varepsilon_1 \varepsilon_2 \gamma_1 + 3 \varepsilon_2^2 \gamma_1 + 
          6 \varepsilon_1 \varepsilon_3 \gamma_1 + 6 \varepsilon_2 \varepsilon_3 \gamma_1 - 6 \varepsilon_3^2 \gamma_1 + 6 \varepsilon_1 \gamma_1^2 + 6 \varepsilon_2 \gamma_1^2 - 12 \varepsilon_3 \gamma_1^2 + 
     \end{equation*}
\begin{equation*}
          16 \gamma_1^3 - 6 \varepsilon_1^2 \gamma_2 + 6 \varepsilon_1 \varepsilon_2 \gamma_2 + 
          3 \varepsilon_2^2 \gamma_2 + 6 \varepsilon_1 \varepsilon_3 \gamma_2 - 12 \varepsilon_2 \varepsilon_3 \gamma_2 +
     \end{equation*}
\begin{equation*}
          3 \varepsilon_3^2 \gamma_2 - 6 \varepsilon_1 \gamma_1 \gamma_2 + 12 \varepsilon_2 \gamma_1 \gamma_2 - 6 \varepsilon_3 \gamma_1 \gamma_2 - 6 \gamma_1^2 \gamma_2 - 12 \varepsilon_1 \gamma_2^2 + 
          6 \varepsilon_2 \gamma_2^2 + 6 \varepsilon_3 \gamma_2^2 - 6 \gamma_1 \gamma_2^2 + 
          16 \gamma_2^3)^2 + 
     \end{equation*}
\begin{equation*}
        4 (-(-\varepsilon_1 - \varepsilon_2 - \varepsilon_3 - 2 \gamma_1 - 2 \gamma_2)^2 + 
           3 (\varepsilon_1 \varepsilon_2 + \varepsilon_1 \varepsilon_3 + \varepsilon_2 \varepsilon_3 + \varepsilon_1 \gamma_1 + 
              \varepsilon_2 \gamma_1 + 2 \varepsilon_3 \gamma_1 + 2 \varepsilon_1 \gamma_2 + \varepsilon_2 \gamma_2 + 
              \varepsilon_3 \gamma_2 + 3 \gamma_1 \gamma_2))^3)^{1/2})^{1/3}
\end{equation*}

\clearpage

\noindent\begin{equation*}
 \omega_3 = 1/3 (\varepsilon_1 + \varepsilon_2 + \varepsilon_3 + 2 \gamma_1 + 
     2 \gamma_2) + ((1 - 
       i \sqrt{3}) (-(-\varepsilon_1 - \varepsilon_2 - \varepsilon_3 - 2 \gamma_1 - 2 \gamma_2)^2 + 
       3 (\varepsilon_1 \varepsilon_2 + \varepsilon_1 \varepsilon_3 + \varepsilon_2 \varepsilon_3 + \varepsilon_1 \gamma_1 +
              \end{equation*}
       \begin{equation*}
        \varepsilon_2 \gamma_1 + 
          2 \varepsilon_3 \gamma_1 + 2 \varepsilon_1 \gamma_2 + \varepsilon_2 \gamma_2 + \varepsilon_3 \gamma_2 + 
          3 \gamma_1 \gamma_2)))/(3\cdot 2^{
     2/3} (2 \varepsilon_1^3 - 3 \varepsilon_1^2 \varepsilon_2 - 3 \varepsilon_1 \varepsilon_2^2 +
            \end{equation*}
       \begin{equation*}
     2 \varepsilon_2^3 - 
       3 \varepsilon_1^2 \varepsilon_3 + 12 \varepsilon_1 \varepsilon_2 \varepsilon_3 - 3 \varepsilon_2^2 \varepsilon_3 - 
       3 \varepsilon_1 \varepsilon_3^2 - 3 \varepsilon_2 \varepsilon_3^2 + 2 \varepsilon_3^3 + 3 \varepsilon_1^2 \gamma_1 -
       12 \varepsilon_1 \varepsilon_2 \gamma_1 + 3 \varepsilon_2^2 \gamma_1 +
              \end{equation*}
       \begin{equation*}
       6 \varepsilon_1 \varepsilon_3 \gamma_1 + 
       6 \varepsilon_2 \varepsilon_3 \gamma_1 - 6 \varepsilon_3^2 \gamma_1 + 6 \varepsilon_1 \gamma_1^2 + 
       6 \varepsilon_2 \gamma_1^2 - 12 \varepsilon_3 \gamma_1^2 + 16 \gamma_1^3 - 6 \varepsilon_1^2 \gamma_2 + 
       6 \varepsilon_1 \varepsilon_2 \gamma_2 +
       3 \varepsilon_2^2 \gamma_2 +
                     \end{equation*}
       \begin{equation*}
        6 \varepsilon_1 \varepsilon_3 \gamma_2 - 
       12 \varepsilon_2 \varepsilon_3 \gamma_2 + 3 \varepsilon_3^2 \gamma_2 - 6 \varepsilon_1 \gamma_1 \gamma_2 + 
       12 \varepsilon_2 \gamma_1 \gamma_2 - 6 \varepsilon_3 \gamma_1 \gamma_2 - 6 \gamma_1^2 \gamma_2 - 
       12 \varepsilon_1 \gamma_2^2 + 6 \varepsilon_2 \gamma_2^2 + 6 \varepsilon_3 \gamma_2^2 - 
                     \end{equation*}
       \begin{equation*}
       6 \gamma_1 \gamma_2^2 + 
       16 \gamma_2^3 + ((2 \varepsilon_1^3 - 3 \varepsilon_1^2 \varepsilon_2 - 
            3 \varepsilon_1 \varepsilon_2^2 + 2 \varepsilon_2^3 - 3 \varepsilon_1^2 \varepsilon_3 + 
            12 \varepsilon_1 \varepsilon_2 \varepsilon_3 - 3 \varepsilon_2^2 \varepsilon_3 - 3 \varepsilon_1 \varepsilon_3^2 -3 \varepsilon_2 \varepsilon_3^2 + 2 \varepsilon_3^3 +
             \end{equation*}
       \begin{equation*}
       3 \varepsilon_1^2 \gamma_1 - 
            12 \varepsilon_1 \varepsilon_2 \gamma_1 + 3 \varepsilon_2^2 \gamma_1 + 6 \varepsilon_1 \varepsilon_3 \gamma_1 + 
            6 \varepsilon_2 \varepsilon_3 \gamma_1 - 6 \varepsilon_3^2 \gamma_1 + 6 \varepsilon_1 \gamma_1^2 + 
            6 \varepsilon_2 \gamma_1^2 - 12 \varepsilon_3 \gamma_1^2 + 16 \gamma_1^3 - 
                         \end{equation*}
       \begin{equation*}
            6 \varepsilon_1^2 \gamma_2 + 6 \varepsilon_1 \varepsilon_2 \gamma_2 + 3 \varepsilon_2^2 \gamma_2 + 
            6 \varepsilon_1 \varepsilon_3 \gamma_2 - 12 \varepsilon_2 \varepsilon_3 \gamma_2 + 3 \varepsilon_3^2 \gamma_2 - 
            6 \varepsilon_1 \gamma_1 \gamma_2 + 12 \varepsilon_2 \gamma_1 \gamma_2 - 6 \varepsilon_3 \gamma_1 \gamma_2 - 
            6 \gamma_1^2 \gamma_2 - 12 \varepsilon_1 \gamma_2^2 + 6 \varepsilon_2 \gamma_2^2 + 
                         \end{equation*}
       \begin{equation*}
            6 \varepsilon_3 \gamma_2^2 - 6 \gamma_1 \gamma_2^2 + 16 \gamma_2^3)^2 + 
          4 (-(-\varepsilon_1 - \varepsilon_2 - \varepsilon_3 - 2 \gamma_1 - 2 \gamma_2)^2 + 
             3 (\varepsilon_1 \varepsilon_2 + \varepsilon_1 \varepsilon_3 + \varepsilon_2 \varepsilon_3 + \varepsilon_1 \gamma_1 + 
                \varepsilon_2 \gamma_1 + 2 \varepsilon_3 \gamma_1 + 
                                \end{equation*}
       \begin{equation*}
                2 \varepsilon_1 \gamma_2 + \varepsilon_2 \gamma_2 +              
                \varepsilon_3 \gamma_2 + 3 \gamma_1 \gamma_2))^3)^{1/2})^{1/3}) - (1/(
  6\cdot 2^{1/3}))(1 + I \sqrt{3}) (2 \varepsilon_1^3 - 3 \varepsilon_1^2 \varepsilon_2 - 
     3 \varepsilon_1 \varepsilon_2^2 + 2 \varepsilon_2^3 - 3 \varepsilon_1^2 \varepsilon_3 + 12 \varepsilon_1 \varepsilon_2 \varepsilon_3 - 
                     \end{equation*}
       \begin{equation*}
     3 \varepsilon_2^2 \varepsilon_3 - 3 \varepsilon_1 \varepsilon_3^2 - 3 \varepsilon_2 \varepsilon_3^2 + 2 \varepsilon_3^3 + 
     3 \varepsilon_1^2 \gamma_1 - 12 \varepsilon_1 \varepsilon_2 \gamma_1 + 3 \varepsilon_2^2 \gamma_1 + 
     6 \varepsilon_1 \varepsilon_3 \gamma_1 + 6 \varepsilon_2 \varepsilon_3 \gamma_1 - 6 \varepsilon_3^2 \gamma_1 + 
                     \end{equation*}
       \begin{equation*}
     6 \varepsilon_1 \gamma_1^2 + 6 \varepsilon_2 \gamma_1^2 - 12 \varepsilon_3 \gamma_1^2 + 16 \gamma_1^3 - 
     6 \varepsilon_1^2 \gamma_2 +
     6 \varepsilon_1 \varepsilon_2 \gamma_2 + 3 \varepsilon_2^2 \gamma_2 + 
     6 \varepsilon_1 \varepsilon_3 \gamma_2 -
                     \end{equation*}
       \begin{equation*}
       12 \varepsilon_2 \varepsilon_3 \gamma_2 + 3 \varepsilon_3^2 \gamma_2 - 
     6 \varepsilon_1 \gamma_1 \gamma_2 + 12 \varepsilon_2 \gamma_1 \gamma_2 - 6 \varepsilon_3 \gamma_1 \gamma_2 - 
     6 \gamma_1^2 \gamma_2 - 12 \varepsilon_1 \gamma_2^2 + 6 \varepsilon_2 \gamma_2^2 + 6 \varepsilon_3 \gamma_2^2 - 
     6 \gamma_1 \gamma_2^2 + 
                     \end{equation*}
       \begin{equation*}
     16 \gamma_2^3 + ((2 \varepsilon_1^3 - 3 \varepsilon_1^2 \varepsilon_2 - 3 \varepsilon_1 \varepsilon_2^2 + 
          2 \varepsilon_2^3 - 3 \varepsilon_1^2 \varepsilon_3 + 12 \varepsilon_1 \varepsilon_2 \varepsilon_3 - 
          3 \varepsilon_2^2 \varepsilon_3 - 3 \varepsilon_1 \varepsilon_3^2 - 3 \varepsilon_2 \varepsilon_3^2 + 2 \varepsilon_3^3 + 
                          \end{equation*}
       \begin{equation*}
          3 \varepsilon_1^2 \gamma_1 - 12 \varepsilon_1 \varepsilon_2 \gamma_1 + 3 \varepsilon_2^2 \gamma_1 + 
          6 \varepsilon_1 \varepsilon_3 \gamma_1 + 6 \varepsilon_2 \varepsilon_3 \gamma_1 - 6 \varepsilon_3^2 \gamma_1 + 
          6 \varepsilon_1 \gamma_1^2 + 6 \varepsilon_2 \gamma_1^2 - 12 \varepsilon_3 \gamma_1^2 + 
          16 \gamma_1^3 - 6 \varepsilon_1^2 \gamma_2 + 6 \varepsilon_1 \varepsilon_2 \gamma_2 + 
                          \end{equation*}
       \begin{equation*}
          3 \varepsilon_2^2 \gamma_2 + 6 \varepsilon_1 \varepsilon_3 \gamma_2 - 12 \varepsilon_2 \varepsilon_3 \gamma_2 + 
          3 \varepsilon_3^2 \gamma_2 - 6 \varepsilon_1 \gamma_1 \gamma_2 + 12 \varepsilon_2 \gamma_1 \gamma_2 - 
          6 \varepsilon_3 \gamma_1 \gamma_2 - 6 \gamma_1^2 \gamma_2 - 12 \varepsilon_1 \gamma_2^2 + 
          6 \varepsilon_2 \gamma_2^2 + 6 \varepsilon_3 \gamma_2^2 - 6 \gamma_1 \gamma_2^2 + 
                          \end{equation*}
       \begin{equation*}
          16 \gamma_2^3)^2 + 
        4 (-(-\varepsilon_1 - \varepsilon_2 - \varepsilon_3 - 2 \gamma_1 - 2 \gamma_2)^2 + 
           3 (\varepsilon_1 \varepsilon_2 + \varepsilon_1 \varepsilon_3 + \varepsilon_2 \varepsilon_3 + \varepsilon_1 \gamma_1 + 
              \varepsilon_2 \gamma_1 + 2 \varepsilon_3 \gamma_1 + 2 \varepsilon_1 \gamma_2 + \varepsilon_2 \gamma_2 + 
                              \end{equation*}
       \begin{equation*}
              \varepsilon_3 \gamma_2 + 3 \gamma_1 \gamma_2))^3)^{1/2})^{1/3}
\end{equation*}

\normalsize

\clearpage

\section{Appendix: Expressions for $\omega_j$ in the $N=4$, $N=5$ and $N=6$ cases, with a single $T_c$ and Josephson coupling}\label{Ap2}
In the main text of this article we have focused on the $N=2$ and $N=3$ cases, because they are analytically solvable and relatively maneagable. Here, let us brielfy comment on the $N>3$ cases. Their main difficulty is to solve the eigenvalue problem of the corresponding $N\times N$ matrix, and associated $N^{\rm th}$ order polynomical equation. In general this is not feasible for $N>3$. However, we found that in the case $\varepsilon_1=\varepsilon_2=\dots \varepsilon_N$   and $\gamma_1=\gamma_2=\dots \gamma_{N-1}$ (\ie, a single $T_c$ and Josephson coupling) it is possible to rewrite the $N=4, 5$ and $6$ polynomials in a solvable form (we were unable to solve the $N=7$ case). We provide those solutions in this Appendix.

For $N=4$ and $\varepsilon_1=\varepsilon_2=\varepsilon_3=\varepsilon_4 (=\varepsilon)$   and $\gamma_1=\gamma_2=\gamma_3=\gamma_4 (=\gamma)$, we found:

\begin{equation}
    \omega_1 = \varepsilon
\end{equation}
\begin{equation}
    \omega_2 = \varepsilon  + 2\gamma
\end{equation}
\begin{equation}
    \omega_3 = \varepsilon  + 2\gamma -\sqrt{2}\gamma
\end{equation}
\begin{equation}
    \omega_4 = \varepsilon  + 2\gamma +\sqrt{2}\gamma
\end{equation}

For $N=5$ and $\varepsilon_1=\varepsilon_2=\dots \varepsilon_5 (=\varepsilon)$   and $\gamma_1=\gamma_2=\dots \gamma_5 (=\gamma)$, we found:
\begin{equation}
    \omega_1 = \varepsilon
\end{equation}
\begin{equation}
    \omega_2 = \frac{1}{2} ( 2\varepsilon  + 3\gamma -\sqrt{5}\gamma )
\end{equation}
\begin{equation}
    \omega_3 = \frac{1}{2} ( 2\varepsilon  + 5\gamma -\sqrt{5}\gamma )
\end{equation}
\begin{equation}
    \omega_4 = \frac{1}{2} ( 2\varepsilon  + 3\gamma +\sqrt{5}\gamma )
\end{equation}
\begin{equation}
    \omega_5 = \frac{1}{2} ( 2\varepsilon  + 5\gamma +\sqrt{5}\gamma )
\end{equation}

\clearpage

For $N=6$ and $\varepsilon_1=\varepsilon_2=\dots \varepsilon_6 (=\varepsilon)$   and $\gamma_1=\gamma_2=\dots \gamma_6 (=\gamma)$, we found:

\begin{equation}
    \omega_1 = \varepsilon
\end{equation}
\begin{equation}
    \omega_2 = \varepsilon + \gamma
\end{equation}
\begin{equation}
    \omega_3 = \varepsilon + 2\gamma
\end{equation}
\begin{equation}
    \omega_4 = \varepsilon + 3\gamma
\end{equation}
\begin{equation}
    \omega_5 = \varepsilon + 2\gamma - \sqrt{3}\gamma
\end{equation}
\begin{equation}
    \omega_6 = \varepsilon + 2\gamma + \sqrt{3}\gamma
\end{equation}

\clearpage

\section{Supplementary information}

Here we provide a figure exemplifying the results for the superconducting fluctuations when considering a different critical temperature for each plane, to show that, as it could be expected, the main effect becomes the simple shift of the effective $T_c$ of the fluctuations, overshadowing the dimensional crossover effects to which our paper is mainly devoted. In particular, we take as example $N=2$ and $T_{c1}=70$K, $T_{c2}=80$K. As can be seen in the figure, when $\gamma$ is small the fluctuations just occur in layer 2 and the fluctuations are the same as if the system was composed of only that layer with upper $T_c$. As $\gamma$ increases though, any fluctuation in layer 2 has to produce an effect in the layer 1 that has a lower critical temperature, and for sufficiently large $\gamma$ the main contribution to the total energy cost of the fluctuations comes from layer 1. Thus, for large $\gamma$ fluctuations simply behave as the ones of a single layer referred to the lower $T_c$, as expected.

\begin{figure}[h]
\begin{center}
\includegraphics[width=0.5\textwidth]{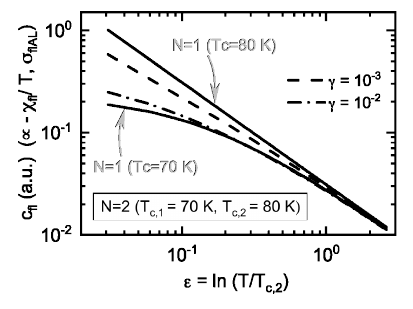}\end{center}
\footnotesize {\bf Fig. SI-1~} Dashed lines: Fluctuation specific heat $c_{\rm fl}$ from our expressions for two-layer superconductors with different critical temperatures $T_{c1}=70$K and $T_{c2}=80$K, for various values of the  Josephson coupling between layers $\gamma$, as a function of the reduced temperature $\varepsilon=\ln(T/T_{c2})$. Continuous lines: The results using $N=1$ and the upper or lower $T_c$, that also correspond to the results using $N=2$ and $\gamma=0$ or $\gamma=\infty$, respectively.
\end{figure}


\begin{thebibliography}{99}





\bibitem{bollinger} Bollinger AT, Bo\v{z}ovi\'c I (2016) Two-dimensional superconductivity in the cuprates revealed by atomic-layer-by-layer molecular beam epitaxy. Supercond Sci Technol 29:103001. https://doi.org/10.1088/0953-2048/29/10/103001 and the rest of articles in the same special issue. 


\bibitem{samples1} Sen K, Marsik P, Das S, Perret E, de Andres Prada R, Alberca A, Bi\v{s}kup N, Varela M,  Bernhard C (2017) Superconductivity and charge-carrier localization in ultrathin La$_{1.85}$Sr$_{0.15}$CuO$_4 $/La$_2$CuO$_4$ bilayers. Phys Rev B 95:214506. https://doi.org/10.1103/PhysRevB.95.214506. 

\bibitem{alegria}   Alegria LD, B{\o}ttcher CGL, Saydjari AK, Pierce AT, Lee SH, Harvey SP, Vool U, Yacoby A (2021) High-energy quasiparticle injection into mesoscopic superconductors.   Nat Nanotech 16:404-408. https://doi.org/10.1038/s41565-020-00834-8. 




\bibitem{katzer} Katzer C, Stahl C, Michalowski P, Treiber S, Schmidl F, Seidel P, Albrecht J, Sch\"utz G (2013) Gold nanocrystals in high-temperature superconducting films:  creation of pinning patterns of choice. New J Phys 15:113029. https://doi.org/10.1088/1367-2630/15/11/113029. 



\bibitem{layer-Fe} Wang F, Lee DH (2011) The Electron-Pairing Mechanism of Iron-Based Superconductors. Science 332:200-204. https://doi.org/10.1126/science.1200182. 

\bibitem{ramallovidal} Ramallo MV, Vidal F (1999) Fluctuation specific heat in multilayered superconductors: Bilayered Gaussian-Ginzburg-Landau scenario for the thermal fluctuations of Cooper pairs around T$_c$ in YBa$_2$Cu$_3$O $_{7-\delta}$. Phys Rev B 59:4475-4485. https://doi.org/10.1103/PhysRevB.59.4475. 





\bibitem{hohenberg} Hohenberg PC, Halperin BI (1977) Theory of dynamic critical phenomena. Rev Mod Phys 49:435-479. https://doi.org/10.1103/RevModPhys.49.435.



\bibitem{KT} Kosterlitz JM, Thouless D (1973) Ordering, metastability and phase transitions in two-dimensional systems. J Phys C 6:1181-1203. https://doi.org/10.1088/0022-3719/6/7/010. 

\bibitem{HN} Halperin BI, Nelson DR (1979) Resistive transition in superconducting films. J Low Temp  Phys 36:599-616. https://doi.org/10.1007/BF00116988. 

\bibitem{LD}  Lawrence WE, Doniach S (1971) Theory of Layer Structure Superconductors. In Kanda E (ed) Proc. Twelfth International Conference on Low Temperature Physics. Academic Press of Japan, Tokio, pp 361-362.

\bibitem{vina} Vi\~na J,  Camp\'a JA, Carballeira C,  Curr\'as SR, Maignan A, Ramallo MV, Rasines I, Veira JA, Wagner P, Vidal F (2002) Universal behavior of the in-plane paraconductivity of cuprate superconductors in the short-wavelength fluctuation regime. Phys Rev B 65:212509. https://doi.org/10.1103/PhysRevB.65.212509. 

\bibitem{carballeira} Carballeira C, Mosqueira J, Ramallo MV, Veira JA, Vidal F (2001) Fluctuation-induced diamagnetism in bulk isotropic superconductors at high reduced temperatures.  J of Physics: Cond Matter 13:9271-9279. https://doi.org/10.1088/0953-8984/13/41/316. 


\bibitem{rey} Rey RI, Ramos-\'Alvarez A, Mosqueira J, Ramallo MV, Vidal F (2013) Comment on ``Diamagnetism and Cooper pairing above T$_c$ in cuprates''. Phys Rev B 87:056501. https://doi.org/10.1103/PhysRevB.87.056501. 








\bibitem{tsuzuki} Tsuzuki T (1972) On the long-range order in superconducting intercalated layer compounds. J Low Temp Phys 9:525-538. https://doi.org/10.1007/BF00655310. 

\bibitem{quader}  Quader KF, Abrahams E (1988) Superconducting fluctuations in specific heat in a magnetic field: dimensional crossover. Phys Rev B 38:11977-11980. https://doi.org/10.1103/PhysRevB.38.11977.






\bibitem{thouless} Thouless DJ (1960) Perturbation theory in statistical mechanics and the theory of superconductivity. Annals of Physics 10:553-588. https://doi.org/10.1016/0003-4916(60)90122-6. 



\bibitem{ferrell} Ferrell RA (1969) Fluctuations and the superconducting phase transition: Critical specific heat and paraconductivity. J Low Temp Phys 1:241-271. https://doi.org/10.1007/BF00628412. 





\bibitem{dorsey} Ullah S, Dorsey AT (1991) Effect of fluctuations on the transport properties of type-II superconductors in a magnetic field. Phys Rev B 44:262-273. https://doi.org/10.1103/PhysRevB.44.262. 


\bibitem{lang1}Puica I, Lang W (2003) Critical fluctuation conductivity in layered superconductors in a strong electric field. Phys Rev B 68:054517. https://doi.org/10.1103/PhysRevB.68.054517. 






\bibitem{vidal} Vidal F, Carballeira C, Curr\'as SR, Mosqueira J, Ramallo MV, Veira JA, Vi\~na J (2002) On the consequences of the uncertainty principle on the superconducting fluctuations well inside the normal state.  Europhys Lett 59:754-759. https://doi.org/10.1209/epl/i2002-00190-3. 


\bibitem{prl} Carballeira C, Mosqueira J, Revcolevschi A, Vidal F (2000) First observation for a cuprate superconductor of fluctuation-induced diamagnetism well inside the finite-magnetic-field regime. Phys Rev Lett 84:3157-3160. https://doi.org/10.1103/PhysRevLett.84.3157. 


\bibitem{mosqueira} Mosqueira J, Ramallo MV, Curr\'as SR, Torr\'on C, Vidal F (2002) Fluctuation-induced diamagnetism above the superconducting transition in MgB$_2$. Phys Rev B 65:174522. https://doi.org/10.1103/PhysRevB.65.174522.

\bibitem{klemm} Klemm RA (1990) Phenomenological model of the copper oxide superconductors. Phys Rev B 41:2073-2097. https://doi.org/10.1103/PhysRevB.41.2073. 


\bibitem{buzdin}Baraduc C, Buzdin A (1992) Fluctuations in layered superconductors: different inter-plane coupling and effect on the London penetration depth. Phys Lett A 171:408-414. https://doi.org/10.1016/0375-9601(92)90666-A. 



\bibitem{REF1a} Vargunin A, \"Ord T (2014) Shrinking of the fluctuation region in a two-band superconductor. Superconductor Sci Tech 27:085006-085014. http://dx.doi.org/10.1088/0953-2048/27/8/085006. 


\bibitem{REF1b} Galteland PN, Subd\o A (2016) Current loops, phase transitions, and the Higgs mechanism in Josephson-coupled multicomponent superconductors. Phys Rev B 94:054518. https://doi.org/10.1103/PhysRevB.94.054518. 



\bibitem{REF1c} Sellin K, Babaev E (2016) First-order phase transition and tricritical point in multiband U(1)  London superconductors. Phys Rev B 93: 054524. https://doi.org/10.1103/PhysRevB.93.054524.

\bibitem{REF4a} Stanev V, Te\v{s}anovi\'c Z (2010) Three-band superconductivity and the order parameter that breaks time-reversal symmetry. Phys Rev B 81:134522. https://doi.org/10.1103/PhysRevB.81.134522. 



\bibitem{LD2} Chapman SJ, Du Q, Gunzburger MD (1995) On the Lawrence–Doniach and Anisotropic Ginzburg–Landau Models for Layered Superconductors. SIAM J Appl Phys 55:156-174. https://doi.org/10.1137/S0036139993256837. 




































\end{thebibliography}
\end{document}